\documentclass[trackchanges,twocolumn]{aastex701}

\usepackage{amsmath}

\newcommand{\lya}{Ly$\alpha$ }
\newcommand{\zre}{$z_\mathrm{re}$}
\newcommand{\zobs}{$z_\mathrm{obs}$}

\newcommand{\HI}{H\,I}
\newcommand{\HeII}{He\,II}

\begin{document}

\title{Unveiling the dark matter nature with reionization relics}

\author[orcid=0009-0002-4542-2621,sname=Zhang,gname=Yao]{Yao Zhang}
\altaffiliation{These authors contributed equally to this work.}
\affiliation{Department of Astronomy, Tsinghua University, Beijing, 100084, People’s Republic of China}
\email{zhangyao22@mails.tsinghua.edu.cn}  

\author[orcid=0000-0002-6998-6678]{Paulo Montero-Camacho}
\altaffiliation{These authors contributed equally to this work.}
\affiliation{Department of Mathematics and Theory, Peng Cheng Laboratory, Shenzhen, 518066, Guangdong, People’s Republic of China}
\affiliation{Department of Astronomy, Tsinghua University, Beijing, 100084, People’s Republic of China}
\email[show]{pmontero@pcl.ac.cn}

\author[gname=Catalina,sname=Morales-Gutiérrez]{Catalina Morales-Gutiérrez}
\affiliation{Department of Physics and Astronomy, University of Waterloo, Waterloo, ON N2L 3G1, Ontario, Canada}
\affiliation{Department of Physics, University of Costa Rica, 11501, San José, Costa Rica}
\affiliation{Space Research Center (CINESPA), University of Costa Rica, 11501, San José, Costa Rica}
\email{c3morale@uwaterloo.ca}

\author[0000-0001-8290-5417]{Heyang Long}
\affiliation{Department of Physics, The Ohio State University, Columbus, 43210, Ohio, USA}
\affiliation{Center for Cosmology and Astroparticle Physics, The Ohio State University, Columbus, 43210, Ohio, USA}
\email{hylong1024@gmail.com}

\author[0000-0002-2951-4932]{Christopher M. Hirata}
\affiliation{Department of Physics, The Ohio State University, Columbus, 43210, Ohio, USA}
\affiliation{Center for Cosmology and Astroparticle Physics, The Ohio State University, Columbus, 43210, Ohio, USA}
\affiliation{Department of Astronomy, The Ohio State University, Columbus, 43210, Ohio, USA}
\email{hirata.10@osu.edu}

\author[0000-0002-1301-3893]{Yi Mao}
\affiliation{Department of Astronomy, Tsinghua University, Beijing, 100084, People’s Republic of China}
\email[show]{ymao@tsinghua.edu.cn}

\begin{abstract}
Dark matter constitutes roughly one-fourth of the Universe, yet its physical nature remains unknown. Warm dark matter (WDM), a class of dark matter candidates, has non-negligible velocity dispersion that suppresses the formation of small-scale cosmic structures. Current constraints therefore rely mainly on small-scale probes such as the Lyman-alpha (Ly$\alpha$) forest and Milky Way observations of satellite galaxies and stellar streams. We propose a novel large-scale probe based on long-lived ``reionization relics'': because the thermal and dynamical evolution of the intergalactic medium depends on the \emph{local} reionization redshift, patchy reionization imprints additional large-scale fluctuations in Ly$\alpha$ forest opacity and post-reionization \HI\ traced by 21\,cm intensity mapping. The strength of these imprints depends on WDM through both small-scale gas evolution and WDM-driven changes in the reionization history. For example, the Ly$\alpha$ (21\,cm) power spectrum in 3\,keV WDM differs from cold dark matter by $\simeq 19\%$ ($\simeq 19\%$) at $k=0.05~\mathrm{Mpc^{-1}}$ at $z=4$ ($z=5.5$) when reionization relics are included. Using Ly$\alpha$ forest with a covariance model designed to mimic the capabilities of the Dark Energy Spectroscopic Instrument (DESI), we forecast a constraint of $m_\mathrm{WDM}>5.0$~keV (95\%), which improves to $m_\mathrm{WDM}>7.1$~keV when combined with 21~cm intensity-mapping observations from the Square Kilometre Array (SKA). The next-generation surveys can further strengthen the current best lower bounds from 9.7 to 39~keV.
\end{abstract}


\section{Introduction} 
The $\Lambda$-Cold Dark Matter ($\Lambda$CDM) framework successfully explains large-scale phenomena like the cosmic microwave background \citep[e.g.,][]{2020A&A...641A...6P}, but has historically faced tensions on small scales (see \citet{2017ARA&A..55..343B} for a review). Warm dark matter (WDM) suppresses structures below the free-streaming scale through its residual thermal velocities and could alleviate these tensions. Alternative approaches -- including completeness correction \citep[e.g.,][]{2018PhRvL.121u1302K} and baryonic feedback \citep[e.g.,][]{2016MNRAS.457.1931S} -- also address these discrepancies. Nevertheless, if dark matter has particle origins, investigating its ``warmth'' probes its properties and formation, as CDM represents the zero-thermal-velocity limit of WDM. Many efforts have been devoted to constraining the mass of WDM particles ($m_\mathrm{WDM}$). The Lyman-alpha (Ly$\alpha$) forest -- absorption features in quasar spectra from neutral hydrogen (\HI) along the line of sight (LoS) -- is highly sensitive to small-scale structure. Milky way observations of satellite galaxies and stellar streams \citep[e.g.,][]{2021ApJ...917....7N,2021JCAP...10..043B} and 21 cm intensity mapping (IM) \citep[e.g.,][]{2014MNRAS.438.2664S, 2015JCAP...07..047C} arising from the hyperfine splitting of the ground state of hydrogen also constrain $m_\mathrm{WDM}$. The strongest current limit, $m_\mathrm{WDM}>9.7~\mathrm{keV}$ \citep{2021ApJ...917....7N}, combines strong gravitational lenses and Milky Way satellite galaxies. Figure~\ref{fig:mass_wdm_constraints} illustrates representative constraints (see Appendix \ref{App:WDM} for details).

\begin{figure*}[htb!]
\centering
\includegraphics[width=1.0\textwidth]{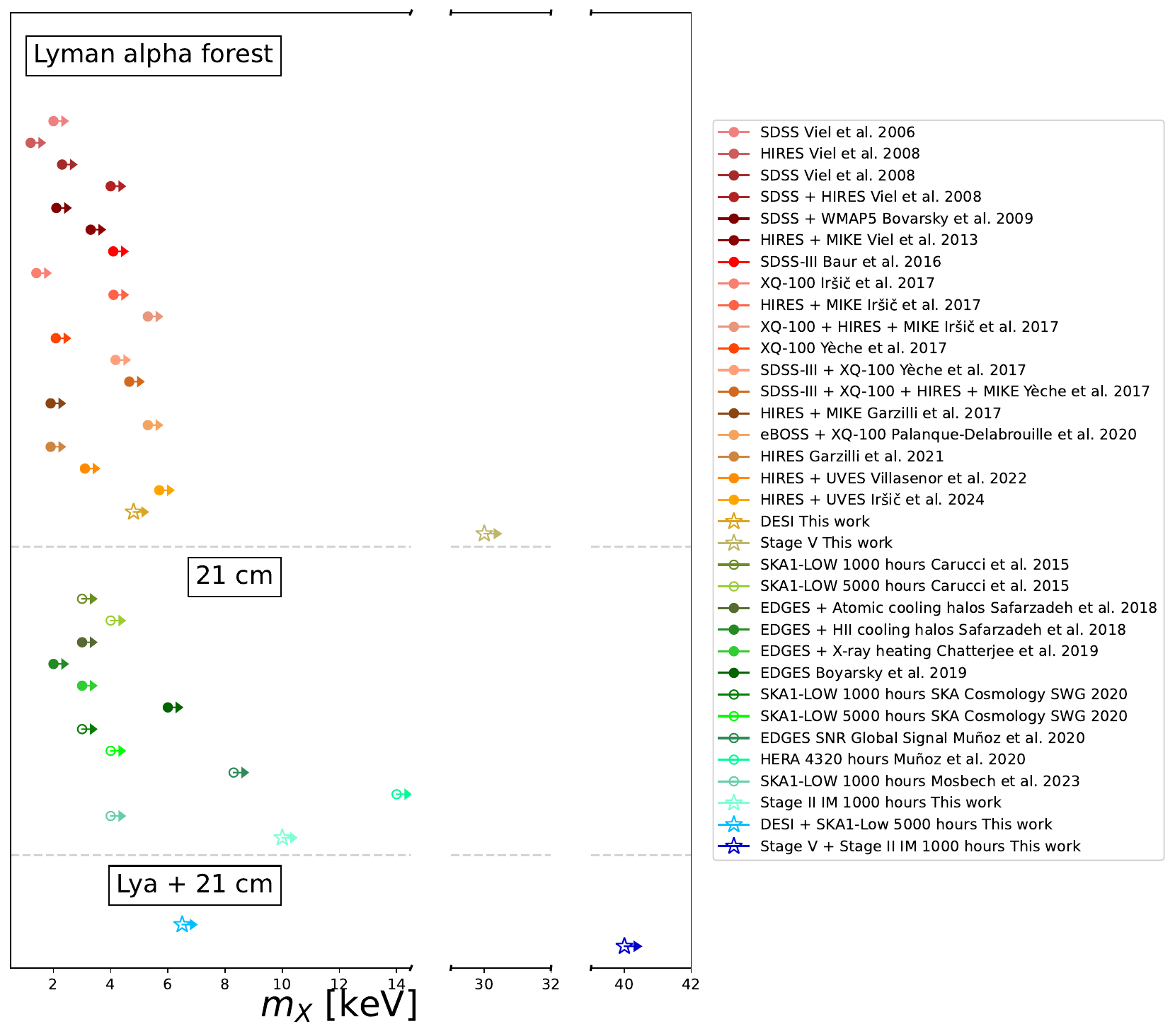}
\caption{WDM mass constraints. Insights from Ly$\alpha$ forest and 21~cm IM. The dot (star) with an arrow represents the lower bound of the WDM mass obtained in the literature (in this work). The data include both observed (filled markers) and forecast (unfilled markers) values from surveys in the Ly$\alpha$ forest, 21~cm IM, and their combination, providing a comprehensive overview of WDM mass constraints across a range of observational techniques (see Appendix~\ref{App:WDM} for details).}
\label{fig:mass_wdm_constraints}
\end{figure*}

While current constraints on WDM mass mainly leverage small-scale observations ($k\gtrsim1~\mathrm{Mpc^{-1}}$), this work proposes a novel large-scale constraint mechanism via the ``memory of reionization'' effect. This effect links the timing of \emph{local} reionization ($z_\mathrm{re}$) to the changes in the transparency to \lya photons of the intergalactic medium (IGM) and \HI\ density.

Reionization raises gas temperatures to $\sim 2\times10^4$~K, typically followed by cooling, with which the IGM ultimately relaxes to a temperature-density ($T-\Delta$) power-law relation on cosmological timescales \citep{1997MNRAS.292...27H}. The gas in mini-halos and mini-filaments is driven by the increased pressure to expand and ``evaporate'' from shallow potential wells. Meanwhile, the low-density gas surrounding the small structures is compressed to mean density, some heated to $>3\times10^4$~K \citep{2018MNRAS.474.2173H}. For late reionization ($z_\mathrm{re}\lesssim8$), a significant amount of the compressed and heated ``high-entropy mean-density'' (HEMD) gas remains above the median ($T-\Delta$) relation long after reionization (e.g., $z\sim2.5$) \citep{2018MNRAS.474.2173H}.  In the post-reionization era, the IGM is highly ionized, while \HI\ gas mainly resides in halos due to self-shielding. Earlier \emph{local} reionization allows the IGM to cool for longer, leading to a higher recombination rate and increased opacity to \lya photons. Furthermore, earlier local reionization allows more time for pressure to dissipate small-scale baryonic structures and suppress subsequent gas accretion, reducing the baryonic fraction in low-mass halos, therefore lowering \HI\ density \citep{2022MNRAS.513..117L}. 

On large scales, reionization is inhomogeneous. Gas around star-forming galaxies is reionized first; then the ionized bubbles surrounding the ionizing sources gradually expand until they overlap. The spatial variation in $z_\mathrm{re}$ results in fluctuations in the IGM transparency and \HI\ density on ionized-bubble scales, since both quantities are sensitive to $z_\mathrm{re}$. 
Previous works have shown that the imprint of inhomogeneous reionization can suppress 21 cm IM \citep{2023MNRAS.525.6036L} and enhance \lya forest \citep[e.g.,][]{2019MNRAS.487.1047M,2019MNRAS.486.4075O,2019MNRAS.490.3177W,2022MNRAS.509.6119M} power spectra on large scales ($k\lesssim0.4~\mathrm{Mpc^{-1}}$).

In this work, we identify different levels of suppression (enhancement) on 21~cm IM (\lya forest) power spectra across various WDM models. These models predict distinguishable signals on large scales, which can be used to constrain $m_\mathrm{WDM}$. The thermal and dynamical evolution of post-reionization gas differs between WDM models, affecting the sensitivities of IGM transparency and \HI\ density to $z_\mathrm{re}$. Because the WDM suppression of small-scale power reduces the abundance of small-scale structures and the density contrast on small scales, gas expansion in low-mass objects is less prevalent and is less violent for lighter WDM. In addition, the gas has less deviation from the ($T-\Delta$) power-law relation relative to CDM. Moreover, suppressed growth of low-mass halos delays reionization by limiting ionization sources. Toward the end, reionization accelerates in WDM models as larger halos that are unaffected by WDM suppression form and host star formation, and rapidly ionize the IGM.

The rest of this paper is structured as follows. Section \ref{sec:model} describes our models of reionization imprints on the \lya foreset and 21~cm IM power spectra, as well as the hybrid simulations. Section \ref{sec:ps} presents the WDM-dependent enhancement of the \lya power spectrum and suppression of the 21 cm IM power spectrum, along with their physical origins. Section \ref{sec:mcmc} presents MCMC forecasts of WDM constraints from current and near-future surveys, and the impact of reionization astrophysics on the constraints. Section \ref{sec:disscusion} compares our methodology with previous works, discusses modeling challenges, and explores observational prospects using \lya forest and 21 cm IM. Section~\ref{sec:conclusion} summarizes our main conclusions.

\section{Models and simulations}
\label{sec:model}

\subsection{Modeling reionization imprints}
To leading order, the imprint of reionization is modeled as follows \citep{2019MNRAS.487.1047M,2023MNRAS.525.6036L}, in terms of the cross-power spectrum of matter density and $X$, where $X$ is the IGM transparency  for the \lya forest or \HI\ density fluctuation for 21~cm IM.
\begin{equation}
\label{eq:cross_power}
\begin{aligned}
P_{\mathrm{m},X}(k,z_\mathrm{obs})
=&
- \int_{z_\mathrm{min}}^{z_\mathrm{max}}
\frac{\partial X}{\partial z}(z,z_\mathrm{obs}) \\
&\times
P_\mathrm{m,x_{HI}}(k,z)
\frac{D_g(z_\mathrm{obs})}{D_g(z)}
\, dz .
\end{aligned}
\end{equation}
The integration covers the full reionization history. $P_\mathrm{m, x_{HI}}$ is the cross-power spectrum of matter density and neutral hydrogen fraction, capturing their anti-correlation on ionized bubble scales. $D_g$ is the growth factor. We use hybrid simulations to compute $P_{\mathrm{m},X}$, combining high-resolution hydrodynamical simulations to model $X$ with semi-numerical reionization simulations to model $P_{\mathrm{m},X}$ (see Section \ref{sec:hybrid} for details).

\subsection{\lya forest power spectrum}
We model the impact of inhomogeneous reionization on the Ly$\alpha$ transmission by the second term on the right-hand-side (RHS) below \citep{2019MNRAS.487.1047M} 
\begin{equation}
    \label{eq:delta_flux}
    \delta_\mathrm{F}=b_\mathrm{F}(1+\beta_\mathrm{F}\mu ^2)\delta_\mathrm{m}+b_\mathrm{\Gamma}\psi,
\end{equation}
where $b_\mathrm{F}$ is the flux bias, $\beta_\mathrm{F}$ is the redshift space distortion (RSD) parameter, $\delta_\mathrm{m}$ is the matter overdensity, $\mu=\cos{\theta}=k_{\parallel}/k$ is the cosine of the angle to the LoS, and $b_\mathrm{\Gamma}$ is the radiation bias defined by $b_\mathrm{\Gamma}=\partial\ln{\bar F}/\partial \ln{\tau_1}$ \citep{2015JCAP...12..017A, 2018MNRAS.474.2173H} with $\bar{F}$ being the mean transmitted flux. We use $\psi$(\zre, \zobs) to parameterize the transparency of an IGM patch that locally reionized at \zre\ and is observed at \zobs, representing the ``memory of reionization".

The 3D \lya forest power spectrum is given by
\begin{equation}
    \label{eq:P_F}
    P_\mathrm{F}^\mathrm{3D} = b_\mathrm{F}^2 (1 + \beta_\mathrm{F} \mu^2)^2 P_\mathrm{m} + 2 b_\mathrm{F} b_\mathrm{\Gamma} (1 + \beta_\mathrm{F} \mu^2) P_\mathrm{m,\psi}\,,
\end{equation}
where $P_\mathrm{m, \psi}$ is the cross-power spectrum of matter and $\psi$ given by Equation (\ref{eq:cross_power}). The second term on the RHS is the leading-order term of the impact of reionization on the 3D \lya forest power spectrum.

We calculate $\psi$ as follows \citep{2018MNRAS.474.2173H}: at a given $z_\mathrm{obs}$, gas particles in the high-resolution hydrodynamical simulations are assigned a \HI\ abundance proportional to $\Delta^2\alpha_\mathrm{A}(T)$, where $\alpha_\mathrm{A}$ is the case A recombination rate and $\Delta$ is the gas density. Here ``case A" contains two recombination channels: first, recombination directly to the ground state; second, recombination to an excited state, then emitting spectral lines to reach the ground state. Then gas particles are smoothed and \HI\ is interpolated onto a grid, producing a map of optical depth $\tau$ with an arbitrary normalization $\tau_1$. We set the normalization by requiring the effective optical depth to match observations \citep{2007MNRAS.382.1657K} as follows: 
\begin{equation}
    \frac{1}{N} \sum_{i = 1}^{N} \exp{(-\tau_1 \tau_i)} = \exp{(-\tau_\mathrm{eff} (\bar{F}))}\, ,
\end{equation}
at $1.7<z<4$, where $\tau_\mathrm{eff}=0.0023(1+z)^{3.65}$. We use $\tau_1$ to quantify the transparency of gas, and 
\begin{equation}
    \psi(z_\mathrm{re},z_\mathrm{obs}) = \ln{\left[\frac{\tau_1(z_\mathrm{re},z_\mathrm{obs})}{\tau_1(z_\mathrm{re}=8,z_\mathrm{obs},\mathrm{CDM})}\right]}
\end{equation}
to describe the relative transparency of gas reionized at $z_\mathrm{re}$ compared to gas reionized at $z=8$ in CDM. Notably, $\psi>0$ means the former scenario is more transparent because the optical depth cube needs a larger normalization to match the observed $\tau_\mathrm{eff}$. A comparison of the mean transmitted flux $\bar{F}$ we use \citep{2007MNRAS.382.1657K} and more recent works \citep[e.g., ][]{2021MNRAS.506.4389G} shows that the difference between measurements is smaller than 7\% at $2<z<3.8$. We compute the \lya\ forest power spectrum with $\bar{F}$ rescaled by factors of 0.93 and 1.07, and find induced variations of approximately $0.5$--$5$\% at $z = 2$--$4$.

We obtain $b_\mathrm{F}$ and $\beta_\mathrm{F}$ as functions of redshift and $\sigma_8$ by interpolating Table 8 in \citet{2015JCAP...12..017A}. The table covers redshifts from 2.2 to 3.0, but our calculations reach $z=4$. Therefore we introduce a redshift evolution factor \citep{2013A&A...559A..85P}. For $z>3$, we use a pivot redshift $z^* =3$. The redshift evolution of the bias factor is then given by 
\begin{eqnarray}
    \label{eq:bias_redshift_evol}
    b_\mathrm{F}^2(z) [1 + \beta_\mathrm{F}(z) \mu^2]^2 & = & \left(\frac{1+z}{1+z^*}\right)^{3.55}\left(\frac{P_\mathrm{m}(z^*)}{P_\mathrm{m}(z)}\right) \nonumber \\
    && \times \, b_\mathrm{F}^2(z^*)[1+\beta_\mathrm{F}(z^*) \mu^2]^2 \, ,
\end{eqnarray}
where $P_\mathrm{m}$ is the matter power spectrum. We ignore the impact of WDM on $b_\mathrm{F}$ and $\beta_\mathrm{F}$ to save computational resources and isolate the WDM's effect on imprints of reionization from the coefficients' effect on the power spectrum. Besides, the difference of $b_\mathrm{F}$ between WDM and CDM is expected to be negligible on large scales (see Figure~3 of \citet{2005PhRvD..71f3534V}). 

Equation (\ref{eq:P_F}) contains the principal components of our theoretical framework; however, it does not account for important systematic effects inherent to observational \lya analyses, such as continuum distortion, high-column density systems, metal contamination, etc. (for a comprehensive list, see, e.g.,  \citealt{2023JCAP...11..045G}). As a first exploration of the impact of WDM on reionization relics, these corrections lie beyond the scope of this work. Future efforts will incorporate them into a more comprehensive framework. 

\subsection{21\,cm IM power spectrum}
We use $\Xi(z_\mathrm{re}, z_\mathrm{obs})$ to represent the \HI\ density fluctuation induced by inhomogeneous reionization \citep{2023MNRAS.525.6036L}:
\begin{equation}
    \label{eq:delta_HI}
    \delta_\mathrm{HI}=b_\mathrm{HI}(1+\beta_\mathrm{HI} \mu^2 )\delta_\mathrm{m}+\Xi(z_\mathrm{re}, z_\mathrm{obs})\, ,
\end{equation}
where $b_\mathrm{HI}$ is the \HI\ bias as a function of redshift \citep{2018ApJ...866..135V}. $\beta_\mathrm{HI}=f/b_\mathrm{HI}$ is the RSD parameter and $f$ is the linear growth rate which can be calculated analytically \citep{2018JCAP...05..004O}. $\Xi$ is defined as follows:
\begin{equation}
    \label{eq:xi}
    \Xi(z_\mathrm{re}, z_\mathrm{obs})=\ln{\frac{\rho_\mathrm{HI}(z_\mathrm{re}, z_\mathrm{obs})}{\rho_\mathrm{HI}(z_\mathrm{mid}, z_\mathrm{obs})}}\approx\frac{\rho_\mathrm{HI}(z_\mathrm{re}, z_\mathrm{obs})}{\rho_\mathrm{HI}(z_\mathrm{mid}, z_\mathrm{obs})}-1\, ,
\end{equation}
where $z_\mathrm{mid}$ is the midpoint of reionization and the approximation holds if $\rho_\mathrm{HI}(z_\mathrm{re}, z_\mathrm{obs})/\rho_\mathrm{HI}(z_\mathrm{mid}, z_\mathrm{obs})\approx1$. Thus, $\Xi$ represents the \HI\ density fluctuation induced by patchy reionization compared to uniform $\rho_\mathrm{HI}(z_\mathrm{mid}, z_\mathrm{obs})$ if reionization is homogeneous. 

The 21\,cm IM power spectrum, including the memory of reionization, is given by \citep{2023MNRAS.525.6036L}
\begin{equation}
    \label{eq:P_21}
    P_{21} = \overline{T}^2_\mathrm{b}(b_\mathrm{HI} + \mu^2 f)^2 P_\mathrm{m} + 2 \overline{T}^2_\mathrm{b} (b_\mathrm{HI} + \mu^2 f) P_\mathrm{m,\Xi}\, .
\end{equation}
Here $P_\mathrm{m,\Xi}$ is obtained using Equation (\ref{eq:cross_power}). $\overline{T}_\mathrm{b}$ is the mean brightness temperature \citep{2006ApJ...652..849F}
\begin{equation}
    \overline{T}_\mathrm{b}(z)=27\,\mathrm{mK}\left(\frac{1+z}{10}\right)^{1/2}\left(\frac{\Omega_\mathrm{HI}(z)h^2}{0.023}\right)\left(\frac{0.15}{\Omega_\mathrm{m}h^2}\right)^{1/2}\, ,
\end{equation}
where $\Omega_\mathrm{HI}$ is the cosmological mass density of neutral hydrogen \citep{2015MNRAS.452..217C}.

The \HI\ density is given by
\begin{equation}
\label{eq:rho_HI}
\begin{aligned}
\rho_{\rm HI}(z_{\rm re}, z_{\rm obs})
=&
\int dM_{\rm h}\,
\frac{dn(M_{\rm h}, z_{\rm obs})}
     {dM_{\rm h}} \\
&\times
M_{\rm HI}(M_{\rm h}, z_{\rm obs}, z_{\rm re}) .
\end{aligned}
\end{equation}

Here $dn/dM_{\rm h}$ is the halo mass function. For CDM, we obtain $dn/dM_{\rm h}$ by a fitting formula \citep{2008ApJ...688..709T}. For models involving WDM, we implement the suppression of low-mass halos through the incorporation of a mass-dependent abundance suppression ratio \citep{2022MNRAS.509.1703S}: 
\begin{equation}
    \label{eq:WDM_hmf}
    \frac{n_\mathrm{WDM}(M)}{n_\mathrm{CDM}(M)}=\left(1+\left(a\frac{M_\mathrm{hm}}{M}\right)^b\right)^c\, ,
\end{equation}
where $M_\mathrm{hm}=4\pi\rho_\mathrm{m,0}(\pi/k_\mathrm{hm})^3/3$ and $k_\mathrm{hm}$ is the scale where the WDM transfer function is suppressed by a factor of 2. The coefficients $a,\ b,$ and $c$ can be calculated by the method provided in \citet{2022MNRAS.509.1703S}\footnote{In practice, we use the ``thermal relic" in the examples: \url{https://github.com/jstuecker/ncdm-mass-functions}}. The halo \HI\ mass $M_\mathrm{HI}$ is assumed to be proportional to $M_\mathrm{h}$ with a cutoff at filtering mass $M_\mathrm{F}$ \citep{2000ApJ...542..535G,2022MNRAS.513..117L,2023MNRAS.525.6036L}, below which the baryonic fraction drops below the cosmic mean due to elevated temperature and pressure following reionization. Following \citet{2022MNRAS.513..117L},

\begin{equation}
\begin{aligned}
M_{\rm HI}(M_{\rm h}, z_{\rm obs}, z_{\rm re})
\propto&
\frac{f_b M_{\rm h}}
{\left[
1+(2^{1/3}-1)
{M_F(z_{\rm obs}, z_{\rm re})}/{M_{\rm h}}
\right]^3} .
\end{aligned}
\end{equation}
where $f_\mathrm{b}\equiv\Omega_\mathrm{b}/\Omega_\mathrm{m}$ is the cosmic mean baryonic fraction. $M_\mathrm{F}=4\pi\rho_\mathrm{m}(\pi/k_\mathrm{F})^3/3$ with the filtering scale, $k_\mathrm{F}$, is given by  
\begin{equation}
    k^{-2}_\mathrm{F}=k^{-2}_\mathrm{F,sound}+k^{-2}_\mathrm{F,vbc}\, ,
\end{equation}
where $k^{-2}_\mathrm{F,sound}$ is the contribution from gas pressure (sound speed) and $k^{-2}_\mathrm{F,vbc}$ arises from the streaming velocities between dark matter and baryons at decoupling and its subsequent evolution. $k^{-2}_\mathrm{F,vbc}$ can be calculated analytically while $k^{-2}_\mathrm{F,sound}(t)$ is obtained by an integration over the thermal history
\begin{eqnarray}
    \label{eq:k_sound}
    k^{-2}_\mathrm{F,sound}(t) & = & \int_{t_\mathrm{dec}}^t -\frac{4}{3}t_1^{-2/3}t_\mathrm{dec}^{-1}\frac{\left(t_1^{2/3}-3t_\mathrm{dec}^{2/3}+2t_\mathrm{dec}t_1^{-1/3}\right)}{\left(t^{2/3}-3t_\mathrm{dec}^{2/3}+2t_\mathrm{dec}t^{-1/3}\right)} \nonumber \\ 
    && \times\left(-t_\mathrm{dec}t_1^{-1/3}+t_\mathrm{dec}t^{-1/3}\right)\frac{c_s^2}{(aH)^2}\bigg \rvert _{t_1}dt_1 \, ,
\end{eqnarray}
where $t_\mathrm{dec}$ is the time at decoupling and the sound speed is given by
\begin{equation}
    c_s=\sqrt{\frac{\gamma k_\mathrm{B}T_0}{\mu_\mathrm{m}}}\, .
\end{equation}
$\gamma$ is the index in $T=T_0(\rho/\bar{\rho})^{\gamma-1}$, and $T_0$ is the temperature of mean-density gas. Moreover, $k_\mathrm{B}$ is the Boltzmann constant and $\mu_\mathrm{m}$ is the reduced mass of $\mathrm{H}^+$, $\mathrm{He}^+$ and $e^-$ plasma with 24.5\% He. We obtain $c_s$ from our hydrodynamical simulations. The sound speed is very small before reionization, thus we only integrate from $t_\mathrm{reionization}$ to $t$ in Equation (\ref{eq:k_sound}). Therefore, $k_\mathrm{F,sound},\ M_\mathrm{F}$ and $\rho_\mathrm{HI}$ are functions of \zre\ and \zobs.

\subsection{Hybrid simulations}
\label{sec:hybrid}
\begin{figure*}[htb!]
\centering
\includegraphics[width=1.0\textwidth]{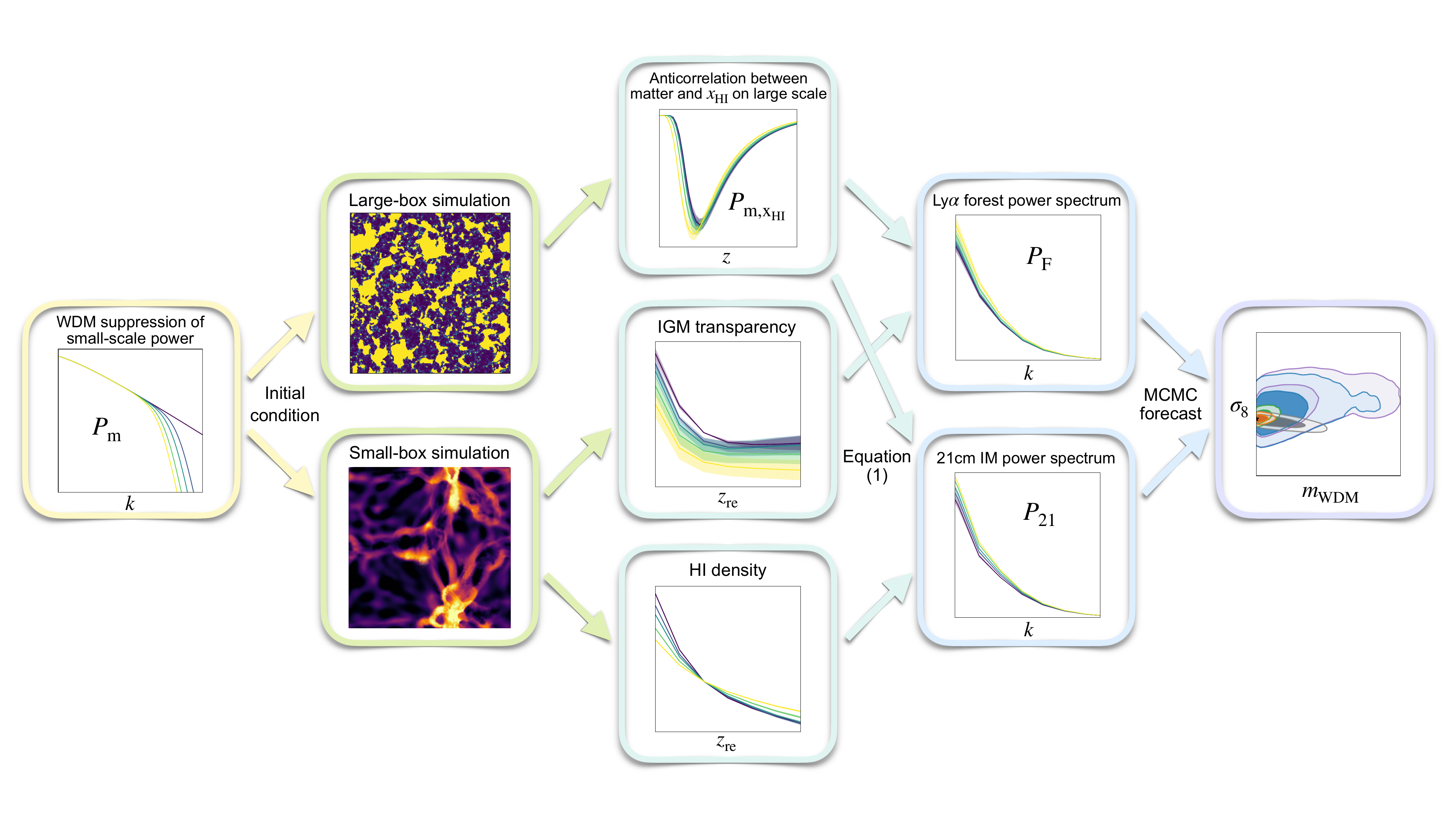}
\caption{Flow diagram illustrating the steps of our hybrid methodology.}
\label{fig:flow}
\end{figure*}

Accurately modeling the imprints of inhomogeneous reionization in the post-reionization era requires a huge dynamical range. First, the simulation must feature sufficient mass resolution to capture masses below the pre-reionization Jeans mass, allowing us to track the response of small-scale structures to the passage of ionization fronts. Inadequate resolution could lead to inaccuracies in modeling the post-reionization IGM and the behavior of minihalos. This limitation arises from the inability to resolve the bi-modality of the $T-\Delta$ relation \citep{2018MNRAS.474.2173H}, i.e., to resolve the HEMD gas, and to probe the smoothing of baryonic structures \citep{2022MNRAS.513..117L, 2023MNRAS.525.6036L}. Second, the patchy nature of reionization must be modeled with enough statistical power to capture its large-scale effects, requiring simulation box lengths of at least $\sim 300$ Mpc \citep{2020MNRAS.495.2354K}. Hence, we adopt a hybrid strategy \citep{2019MNRAS.487.1047M,2023MNRAS.525.6036L}, combining high-mass-resolution hydrodynamical simulations to track small-scale structures and semi-numerical simulations to handle the patchy nature of reionization, as illustrated in Figure~\ref{fig:flow}.

We use a modified version of \textsc{Gadget-2} \citep{2001NewA....6...79S, 2005MNRAS.364.1105S, 2018MNRAS.474.2173H} for the small-scale high-mass resolution simulations. These simulations track the gas evolution after local reionization, and the goals include extracting the transparency of small-scale IGM $\psi(z_\mathrm{re},z_\mathrm{obs})$, determining the sound speed $c_s$ for computing the \HI\ density fluctuation  $\Xi(z_\mathrm{re},z_\mathrm{obs})$, and investigating the variations of $\psi$ and $\Xi$ with \emph{local} reionization redshift $z_\mathrm{re}$. The box length is $1275\, \mathrm{kpc}$, and the number of particles is $2\times192^3$. The particle masses of dark matter and gas are $9.72\times10^3M_\odot$ and $1.81\times10^3M_\odot$, respectively, thus providing sufficient mass resolution to resolve structures below pre-reionization Jeans mass. The convergence tests performed in  \citet{2018MNRAS.474.2173H} demonstrate that for $z_\mathrm{re}=7$ and 9 and $2.5\leq z_\mathrm{obs}\leq4.0$, increasing the mass resolution by a factor of 8 changes $\psi$ by $<7$\%. Similarly, increasing the simulation volume by 8 yields $\psi$ variation at 1$\sigma$ level, except for $z_\mathrm{obs}=4.0$, where deviations are within 10\%. Our configuration was selected to enable sufficient realizations. 

We consider WDM as a thermal relic that was once in thermal equilibrium with photons and later decoupled as the Universe cooled.
WDM was implemented by using the WDM power spectrum as the initial condition of the simulations, excluding thermal velocities in the initial particle velocity distributions. The CDM power spectrum was obtained using {\sc Class} \citep{2011JCAP...07..034B}, with WDM small-scale suppression introduced via a transfer function \citep{2001ApJ...556...93B}
\begin{eqnarray}
    \label{eq:transfer}
    T(k) = [1 + (\alpha k)^{2\nu}]^{-5/\nu} \, ,
\end{eqnarray}
where $\nu = 1.12$ and the scale of the break $\alpha$ is given by \citep{2005PhRvD..71f3534V} 
\begin{equation}
\label{eq:alpha}
\begin{aligned}
\alpha
=\, & 0.049
\left(\frac{m_{\rm WDM}}{1\,{\rm keV}}\right)^{-1.11}
\left(\frac{\Omega_{\rm WDM}}{0.25}\right)^{0.11} \\
& \times
\left(\frac{h}{0.7}\right)^{1.22}
\, h^{-1}\,{\rm Mpc} .
\end{aligned}
\end{equation}
We assume all dark matter is WDM, i.e. $\Omega_\mathrm{WDM}=\Omega_\mathrm{DM}$. Then the matter power spectrum of WDM is
\begin{eqnarray}
    \label{eq:WDM_power}
    P_\mathrm{WDM}(k)=T^2(k)P_\mathrm{CDM}(k) \, .
\end{eqnarray}

All small-box simulations start at $z_\mathrm{dec}=1059$, with a fixed $33~\mathrm{km\,s^{-1}}$ initial streaming velocity between dark matter and baryons \citep{2018MNRAS.474.2173H}. Before reionization, the neutral gas undergoes Compton scattering from the CMB and X-ray preheating \citep{2024MNRAS.529.3666M}. Considering the length of the box is about one cell size of the large-scale semi-numerical simulation, we assume reionization happens at $z_\mathrm{re}$ simultaneously inside the small box ($z_\mathrm{re}$=\{6, 7, 8, 9, 10, 11, 12\}), and gas has a uniform post-reionization temperature of $2\times10^4\,\mathrm{K}$. After reionization, the gas is treated as plasma mainly consisting of $\mathrm{H}^+$, $\mathrm{He^+}$, and $e^-$, and undergoes Compton cooling, recombination cooling, photoionization heating, free-free cooling, and $\mathrm{He^+}$ line cooling, which are implemented in \textsc{Gadget-2} \citep{2018MNRAS.474.2173H}.

We use \textsc{21cmFAST} \citep{2011MNRAS.411..955M,2020JOSS....5.2582M,2019MNRAS.484..933P} for the large-scale simulations to extract reionization history and compute $P_\mathrm{m,x_{HI}}(k,z)$ in Equation \ref{eq:cross_power}, which quantifies the patchiness of reionization on different scales. The box length is 400\, Mpc, with $256^3$ cells for \HI\ and $768^3$ cells for the matter field, and we mostly use \textsc{21cmFAST} default parameters. There are two aspects to include the effect of WDM on the reionization process. First, we compute an effective Jeans mass as a function of $m_\mathrm{WDM}$ \citep{2014MNRAS.438.2664S},
\begin{equation}
\begin{aligned}
M_J \approx\,
& 1.5 \times 10^{10}
\left( \frac{\Omega_{\rm WDM} h^2}{0.15} \right)^{1/2}
\\
& \times
\left( \frac{m_{\rm WDM}}{1\,{\rm keV}} \right)^{-4}
\, M_\odot .
\end{aligned}
\end{equation}
$M_J$ defines a new minimum mass that enters the mean collapse fraction $f_\mathrm{coll}(>M_\mathrm{min},z)$. $M_\mathrm{min} = \mathrm{max}(M_J, M_\mathrm{sf})$ where $M_\mathrm{sf}$ is the minimum halo mass capable of hosting star formation. The mean collapse fraction determines the production rate of photons that heat and ionize the IGM. Second, the cutoff of small-scale matter power spectrum is implemented using the transfer function in Equation (\ref{eq:transfer}). We calculate the comoving cutoff scale $R_c^0$ \citep{2014MNRAS.438.2664S}, at which the power in $k=1/R_c^0$ is reduced by half compared to CDM, then $R_c^0$ is mapped to $\alpha$ -- Equation (\ref{eq:alpha}) -- by $\alpha=0.356R_c^0$ to be consistent with the small-scale simulations.

Throughout this work we assume a background cosmology from the \emph{ Planck} 2015 `TT,TE,EE+\allowbreak lowP+\allowbreak lensing+\allowbreak ext' parameter set \citep{2016A&A...594A..13P}: $\Omega_\mathrm{b}h^2=0.02230,\ \allowbreak \Omega_\mathrm{m}h^2=0.14170,\ \allowbreak H_0=67.74\,\mathrm{km\,s^{-1}\,Mpc^{-1}},\ \allowbreak n_s=0.9667$. For 2-parameter MCMC analysis, we run small-scale and large-scale simulations with 2 parameters forming a grid with 15 points: $m_\mathrm{DM}=\{3\,\mathrm{keV},\ 4\,\mathrm{keV},\ 6\,\mathrm{keV},\ 9\,\mathrm{keV},\ \mathrm{infinity(CDM)}\}$; $\sigma_8=\{0.7659,\ 0.8159,\ 0.8659\}$. For each parameter configuration, we change the random seeds to obtain 4 different realizations. 

\section{Reionization imprints on power spectrum}
\label{sec:ps}
\subsection{WDM-dependent enhancement of \lya power}
\begin{figure*}[htb!]%
\centering
\includegraphics[width=1.0\textwidth]{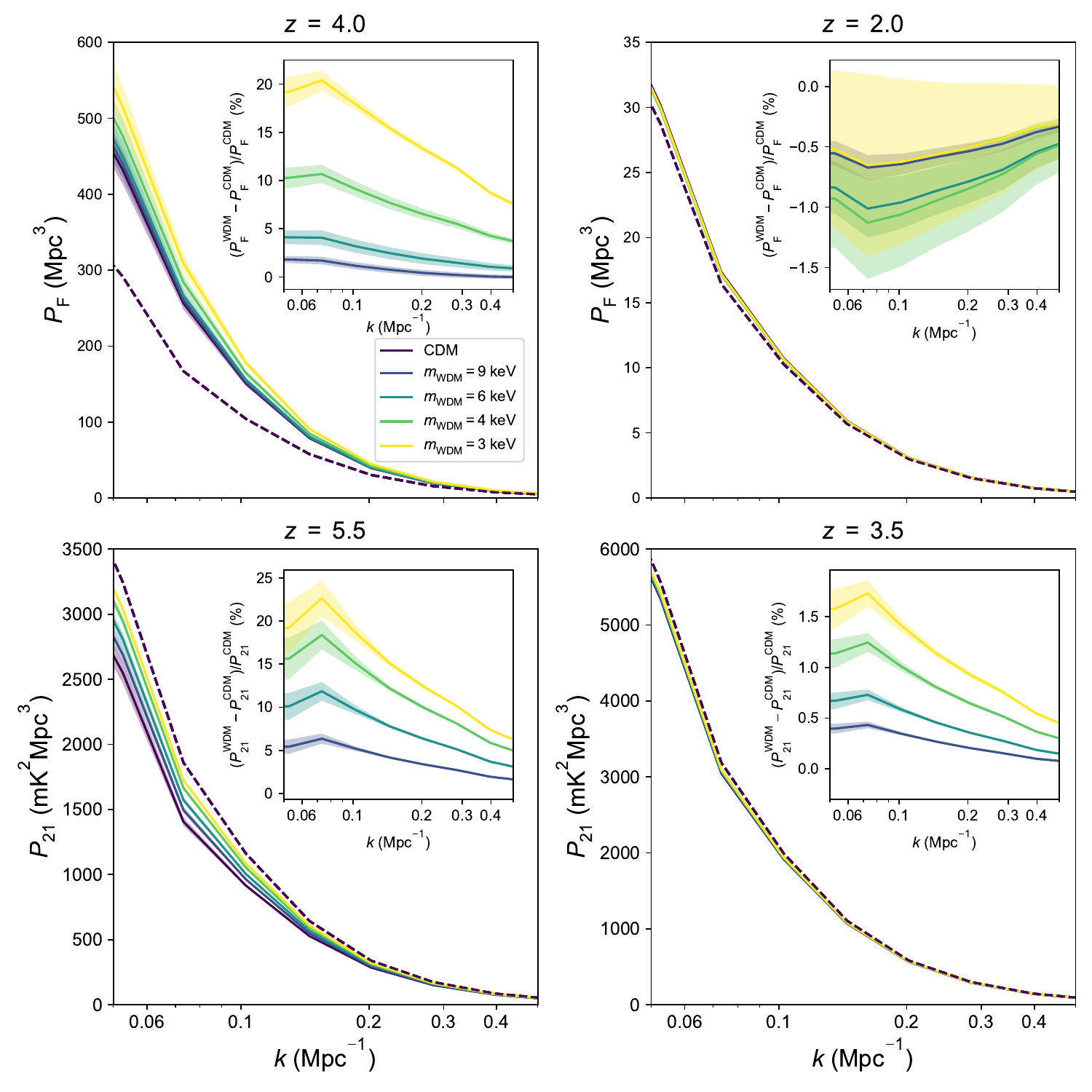}
\caption{The imprint of reionization in different dark matter models.
We show the \lya\ forest power spectrum with $\mu=0.1$ at $z=4$ and $z=2$ in the top row, and the 21~cm IM power spectrum with $\mu=0.9$ at $z=5.5$ and $z=3.5$ in the bottom row, including the imprint of reionization for the different WDM models considered herein (colored solid lines) and for the CDM model (blue solid line). Here, $\mu=\cos{\theta}$ and $\theta$ is the angle between the wavenumber vector $\boldsymbol{k}$ and the LoS. The shaded regions represent the standard deviation of four realizations for each model. For comparison, we show the conventional -- i.e.\ no reionization relics -- linear power spectrum (dashed line) in CDM. The fractional difference of the power spectrum between the WDM and CDM (both including the imprint of reionization) is shown in each inset. }
\label{fig:power_spectrum}
\end{figure*}

Figure~\ref{fig:power_spectrum} shows the 3D \lya and 21~cm IM power spectra for dark-matter models. Reionization enhances large-scale \lya forest power, particularly at higher redshifts and lower $m_\mathrm{WDM}$. At $z_\mathrm{high}^\mathrm{Ly\alpha}=4$, enhancements at $k=0.05\,\mathrm{Mpc^{-1}}$ ($0.2\,\mathrm{Mpc^{-1}}$) reach 75\% (48\%) for 3 keV WDM and 47\% (31\%) for CDM. Including reionization relics, fractional differences between WDM ($m_\mathrm{WDM}=3,\,4,\,6,\,9$ keV) and CDM are 19\% (13\%), 10\% (7\%), 4\% (2\%) and 2\% (0.4\%) at $k=0.05\,\mathrm{Mpc^{-1}}$ ($0.2\,\mathrm{Mpc^{-1}}$), respectively. Without reionization relics, large-scale differences between dark matter models are negligible. At $z_\mathrm{low}^\mathrm{Ly\alpha}=2$, heavier WDM marginally boosts the enhancement ($\sim$4-5\% at $k=0.05\,\mathrm{Mpc^{-1}}$), with a fractional difference within 1.5\% relative to CDM.

The impact of reionization on the \lya forest is an enhancement because without considering reionization, high-density regions are more opaque than low-density regions (note that the flux bias is negative). After considering reionization, high-density regions are even more opaque since they usually reionize earlier and the IGM has more time to cool. The inhomogeneity of reionization enhances the contrast in opacity on scales where \emph{local} $z_\mathrm{re}$ changes, i.e. the ionized bubble scales, which are several to tens of comoving Mpc (cMpc). The degree of large-scale enhancement differs between WDM models due to two factors.

\begin{figure}[htb!]%
\centering
\includegraphics[width=\columnwidth]{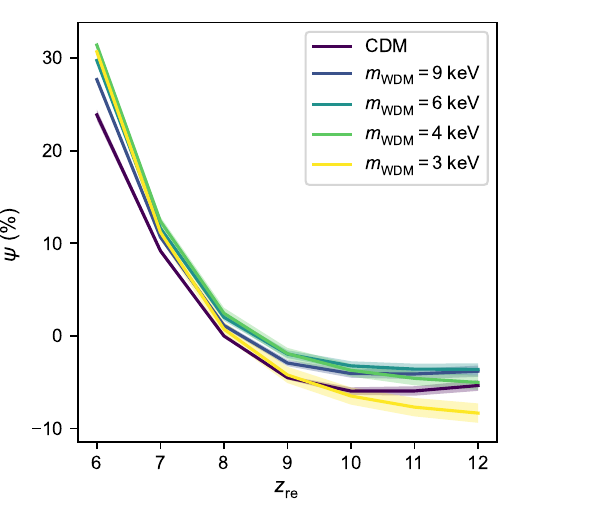}
\caption{Relative transparency of a patch of gas locally reionized at $z_\mathrm{re}$ and observed at $z_\mathrm{obs}=4$ compared to gas reionized at $z_\mathrm{re}=8$ in CDM model. Colored lines represent different dark matter models considered in this work, and the shaded region shows the standard deviation of four realizations.}
\label{fig:transparency}
\end{figure}

\begin{figure*}[htb!]%
\centering
\includegraphics[width=0.8\textwidth]{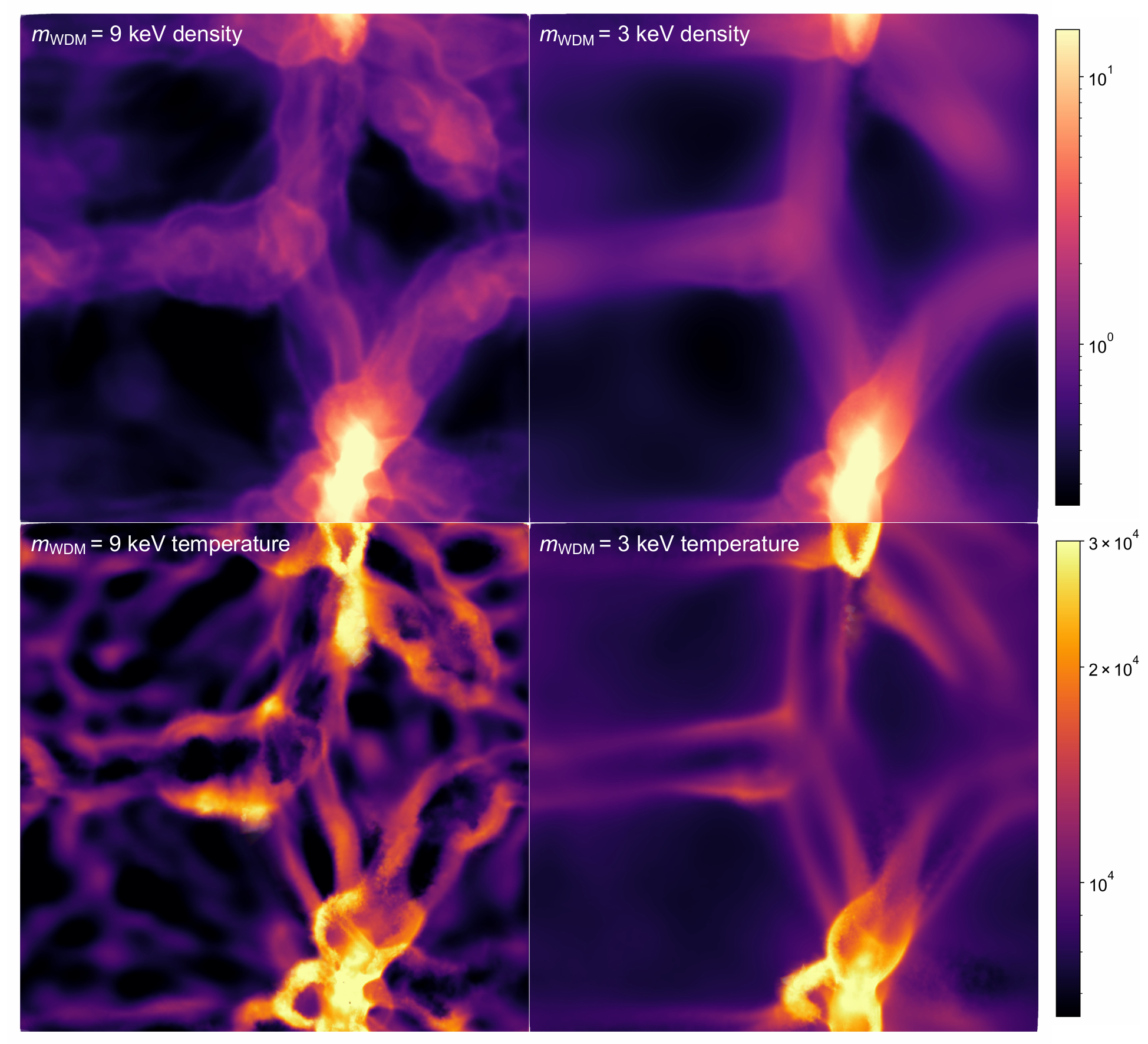}
\caption{The post-reionization ($z_\mathrm{re}=7$) gas overdensity $\rho/\bar{\rho}$ (first row) and temperature in Kelvin (second row) at $z_\mathrm{obs}=5$. The left and right columns correspond to 9~keV and 3~keV WDM models, respectively. The length of each panel is 1275~kpc.}
\label{fig:visualization}
\end{figure*}

\begin{figure*}
\centering
\includegraphics[width=1.0\textwidth]{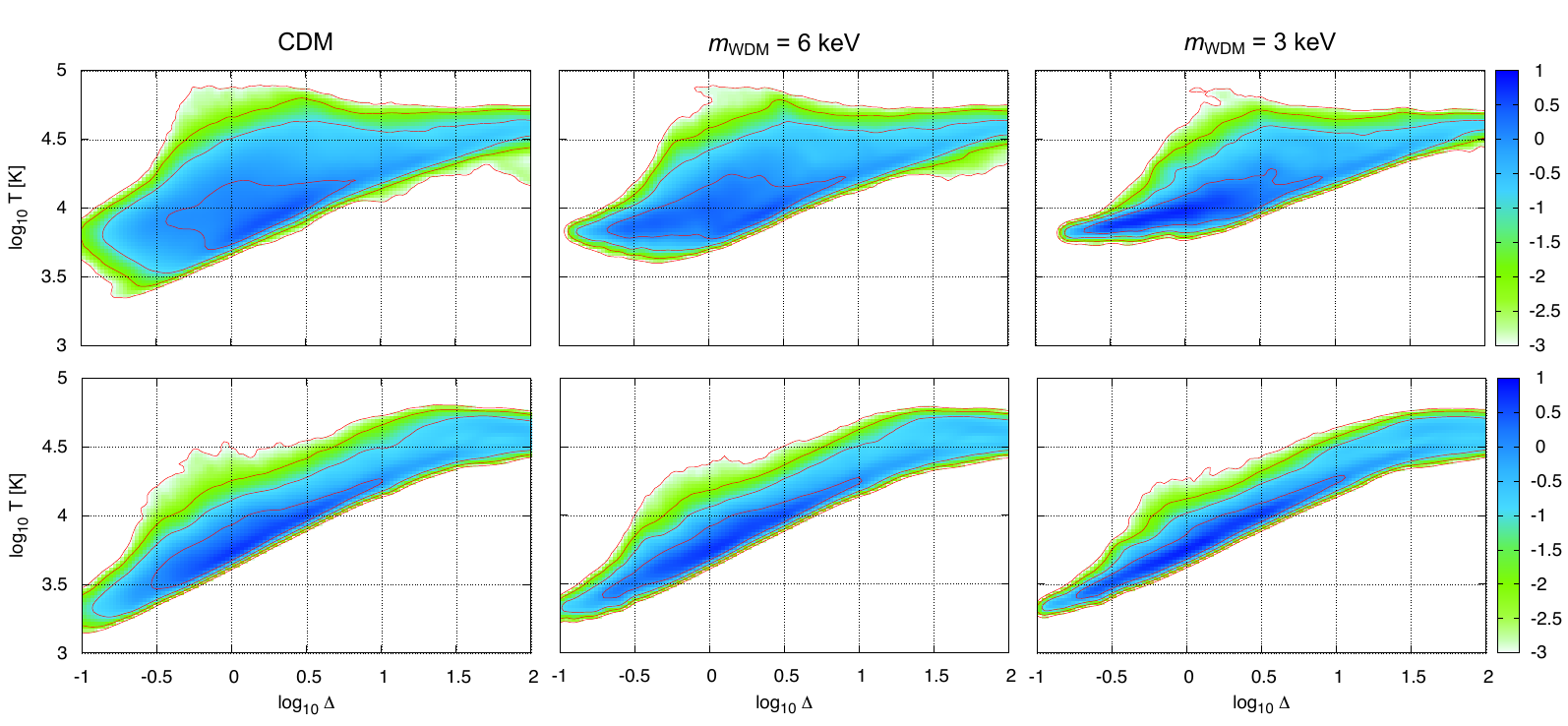}
\caption{The post-reionization temperature-density relation of a patch of gas that locally reionizes at $z_{\rm re}=6$ and is observed at $z_{\rm obs}=4$ (top row) and $z_{\rm obs}=2$ (bottom row). The left, middle, and right columns correspond to CDM, 6 keV WDM and 3 keV WDM, respectively. In each panel, particles are smoothed with a Gaussian kernel density estimation with a full-width at half-maximum of 0.05 dex on each axis. The numbers of the colorbars represent the mass-weighted probability density of the gas $\mathrm{d}P/\mathrm{dlog_{10}}T\mathrm{dlog_{10}}\Delta$ in units of $\mathrm{log_{10}}$. A color of green (-2) means that $\mathrm{d}P/\mathrm{dlog_{10}}T\mathrm{dlog_{10}}\Delta=0.01$. The contours indicate each decade in probability density.}
\label{fig:dinosaur}
\end{figure*}

First, a stronger enhancement occurs when IGM transparency significantly varies with \emph{local} $z_\mathrm{re}$, causing larger fluctuations on large scales. At $\sim z^\mathrm{Ly\alpha}_\mathrm{high}$, this variation is more pronounced for lighter WDM (see Figure \ref{fig:transparency}), as the scarcity of small-scale structures in the voids limits their evaporation and the compression and heating of surrounding low-density gas, leading to less HEMD gas (see Figure~\ref{fig:visualization}). The cooling of low-density IGM is less affected by the dynamical response of the small-scale structures, leading to a more substantial temperature and transparency decrease with increasing $z_\mathrm{re}$. However, at $\sim z^\mathrm{Ly\alpha}_\mathrm{low}$, heavier WDM shows more variation. Long after reionization, transparency sensitivity to $z_\mathrm{re}$ is mainly due to cooling of HEMD gas, which is more abundant and reaches a higher temperature in heavier WDM models (see Figure \ref{fig:dinosaur}), because of a more violent post-reionization dynamical response with more small-scale structures and larger density contrast. These behaviors are captured by the small-box simulations.

\begin{figure}[htb!]%
\centering
\includegraphics[width=\columnwidth]{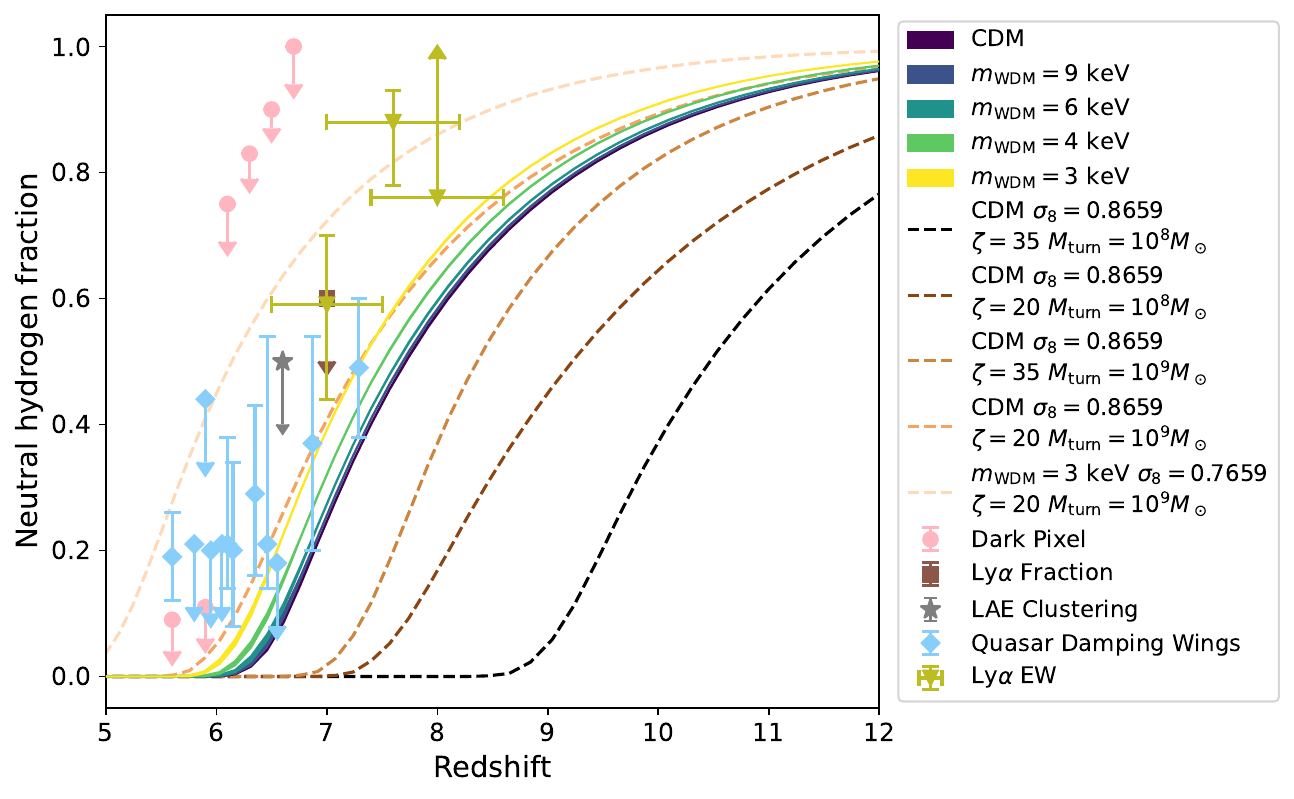}
\caption{Evolution of volume-weighted global neutral hydrogen fraction given by semi-numerical reionization simulations. The solid lines indicate the evolution of $x_\mathrm{HI}$ in different DM models with other parameters set to fiducial values. The dashed lines show five examples when $\sigma_8$, ionizing efficiency $\zeta$ and turnover mass of halos unable to host star-forming galaxies $M_\mathrm{turn}$ change. The rightmost dashed line is the earliest in all 
reionization models considered in this work, and the leftmost is the latest. The data points correspond to observational constraints from various works including dark pixel fractions \citep{2015MNRAS.447..499M,2023ApJ...942...59J}, clustering of early galaxies through their \lya emission \citep{2010ApJ...723..869O,2015MNRAS.453.1843S}, and their \lya equivalent width distribution\citep{2018ApJ...856....2M, 2019ApJ...878...12H, 2019MNRAS.485.3947M}, \lya fraction evolution \citep{2015MNRAS.446..566M}, and quasar damping wings \citep{2022MNRAS.512.5390G,2024MNRAS.530.3208G,2024A&A...688L..26S,2024ApJ...969..162D}.}
\label{fig:global_history}
\end{figure}

Second, the enhancement is greater when the whole reionization process is delayed, which is more pronounced for lighter WDM and captured by the large-box reionization simulations, as shown in Figure~\ref{fig:global_history}. Compared to an early reionization scenario, late reionization results in more regions with lower \emph{local} $z_\mathrm{re}$, i.e. more regions have less time to relax and their transparency is more sensitive to $z_\mathrm{re}$.  

\subsection{WDM-dependent suppression of 21~cm IM power}
For the 21~cm IM, reionization imprints cause large-scale suppression, with stronger effects for heavier WDM. At high redshift, e.g.,\ $z^\mathrm{21\,cm}_\mathrm{high}=5.5$, the suppression at $k=0.05\,\mathrm{Mpc^{-1}}$ ($0.2\,\mathrm{Mpc^{-1}}$) is 21\% (16\%) for CDM and 6\% (5\%) for 3\,keV WDM. The fractional difference between WDM and CDM power spectra at $k=0.05\,\mathrm{Mpc^{-1}}$ ($0.2\,\mathrm{Mpc^{-1}}$) is 19\% (12\%), 15\% (10\%), 10\% (6\%) and 5\% (3\%) 
for $m_\mathrm{WDM}=3,\,4,\,6,\,9$ keV, respectively. At low redshift, e.g.\ $z^\mathrm{21\,cm}_\mathrm{low}=3.5$, the suppression weakens: only approximately 2\% to 4\% at $k=0.05\,\mathrm{Mpc^{-1}}$ for DM models, with fractional difference between WDM and CDM from sub-percent to 1.5\%, depending on WDM mass (9 to 3 keV).

\begin{figure}[htb!]%
\centering
\includegraphics[width=\columnwidth]{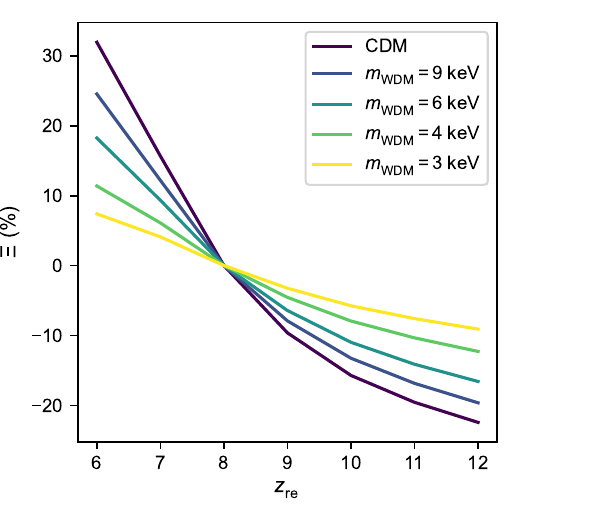}
\caption{Relative \HI\ density of a patch of gas locally reionized at $z_\mathrm{re}$ and observed at $z_\mathrm{obs}=5.5$ compared to gas reionized at $z_\mathrm{re}=8$. Colored lines and shaded regions follow the same convention as in Figure~\ref{fig:transparency}. The shaded regions are too small to be visible.}
\label{fig:rho_HI}
\end{figure}

In the post-reionization era, the 21 cm signal mainly comes from neutral hydrogen in self-shielded halos. The 21\,cm brightness temperature is proportional to the \HI\ density. Without considering reionization, dense regions tend to have high \HI\ density. However, reionization reduces \HI\ density by lowering the baryonic fraction in low-mass halos. When smoothed on ionized bubble scales, the \HI\ density in dense regions is reduced more significantly, leading to smaller \HI\ density fluctuations and suppressed 21\,cm emission fluctuations. In the post-reionization era, cosmological timescales are needed for increased pressure to smooth out small-scale baryonic structures. Dense regions typically reionize earlier and have more time for dynamical response, leading to less \HI\ gas in low-mass halos. We find that the suppression of 21\,cm IM is weaker for lighter WDM at both $ z^\mathrm{21\,cm}_\mathrm{high}$ and $z^\mathrm{21\,cm}_\mathrm{low}$ because of fewer low-mass halos suffering the reduction of baryons by reionization, resulting in a subdued sensitivity of \HI\ density to $z_\mathrm{re}$, as shown in Figure \ref{fig:rho_HI}. Previous works also found large-scale responses to small-scale DM physics (e.g.,  \citealt{2015JCAP...07..047C,2023JCAP...01..002C,2019PhRvD..99b3518C,2020PhRvD.101l3518I}). For instance,  \cite{2015JCAP...07..047C} found a higher amplitude of 21 cm IM power spectrum of WDM compared to CDM without considering reionization relics, because more \HI\ is distributed in the massive halos, which are strongly clustered, due to a deficit of low-mass halos. We note combining this effect with the small-scale sourced reionization relics can further increase the differences among DM models.

\section{MCMC forecasts of WDM constraints}
\label{sec:mcmc}
\subsection{Methods}
With distinguishable signatures on \lya and 21\,cm observations, we use Monte Carlo Markov Chain (MCMC), as implemented in {\textsc emcee} \citep{2013PASP..125..306F}, to predict the constraints on $m_\mathrm{WDM}$ by ongoing and future surveys. Besides $m_\mathrm{WDM}$, we include $\sigma_8$ as a free parameter to allow for degeneracies. The likelihood function is 
\begin{equation}
\label{eq:likelihood}
    \mathcal L\propto \mathrm{exp}(-\frac{1}{2}\sum_\mathrm{bins}(P_\ell(z,\boldsymbol{k})-P_\ell^\mathrm{fid}(z,\boldsymbol{k}))^2/\sigma_{\ell}^2(z,\boldsymbol{k})) \, ,
\end{equation}
where $P_{\ell} = \{P_\mathrm{F}, P_{21} \}$ are given by Equations (\ref{eq:P_F}) and (\ref{eq:P_21}). $P_\ell^\mathrm{fid}$ corresponds to the fiducial model, which is CDM with $\sigma_8=0.8159$ \citep{2016A&A...594A..13P}. $\sigma_{\ell}^2$ represents the covariance of the given observable \citep{2003ApJ...598..720S}\footnote{Note the reduction by a factor of 2 compared to Eq.~(8) in \citep{2003ApJ...598..720S} due to our range in $\mu$ being half of their range.}
\begin{eqnarray}
\label{eq:covariance}
    \sigma_\ell^2(z,\boldsymbol{k})=[P_{\ell,\mathrm{Tot}}^\mathrm{fid}(z,\boldsymbol{k})]^2\frac{4\pi^2}{V_\mathrm{survey}(z)k^2 \Delta k\Delta\mu} ,
\end{eqnarray}
with $\Delta k$ and $\Delta \mu$ the widths of the $k$-bin and $\mu$-bin, respectively. $V_\mathrm{survey}$ is the comoving survey volume. Note that $P_{\ell,\mathrm{Tot}}^\mathrm{fid}$ contains both signal and noise. In practice, we use 1/$m_\mathrm{WDM}$ as a parameter for the MCMC to avoid infinity when representing CDM. We use flat priors for both 1/$m_\mathrm{WDM}$ and $\sigma_8$:
\begin{equation}
\begin{aligned}
p\left(\frac{1}{m_{\rm WDM}}, \sigma_8\right)
=
\begin{cases}
{\rm const.},
& \displaystyle
0 < \frac{1}{m_{\rm WDM}} < \frac{1}{3},
\\[6pt]
&
\displaystyle
0.7659 < \sigma_8 < 0.8659 ,
\\[6pt]
0,
& {\rm otherwise} .
\end{cases}
\end{aligned}
\end{equation}

The total observed \lya forest power spectrum is (see \citealt{2021MNRAS.508.1262M} and references therein)
\begin{equation}
\label{eq:P_F_total}
\begin{aligned}
P_{\rm F,Tot}^{\rm fid}(\boldsymbol{k}, z)
=
&\, P_{\rm F}^{\rm fid}(\boldsymbol{k}, z)
\\
&+
P_{\rm F}^{\rm 1D}(k_{\parallel}, z)
P_{\rm w}^{\rm 2D}(z)
+
P_{\rm N}^{\rm eff}(z) .
\end{aligned}
\end{equation}
The second term represents the aliasing term accounting for the 2D quasar density. $P_\mathrm{F}^\mathrm{1D}$ is the 1D Ly$\alpha$ forest power spectrum along LoS \citep{2013JCAP...05..018F}. $P_\mathrm{w}^\mathrm{2D}$ is the inverse of the effective density of lines of sight, and $P_\mathrm{N}^\mathrm{eff}$ is the weighted noise power spectrum. Both depend on quasar luminosity function \citep{2020RNAAS...4..179Y} and telescope performance \citep{2007PhRvD..76f3009M,2014JCAP...05..023F,2021MNRAS.508.1262M}.

We consider a DESI-like 5-year 3D flux survey based on the expected performance of the Dark Energy Spectroscopic Instrument (DESI) \citep{2022AJ....164..207D,2023JCAP...11..045G,2024MNRAS.528.6666R}, as well as Stage-V spectroscopic surveys such as MUltiplexed Survey Telescope (MUST) \citep{2024arXiv241107970Z}. The DESI spectroscopic pipeline is described in \citet{2023AJ....165..144G}. We assume that Stage-V surveys will cover twice DESI's sky coverage (28,000 $\mathrm{deg^2}$), which is possible if we combine different surveys, and have one-third of DESI's noise.
The forecasts use two redshift bins ($z=\{2.5,3.5\}$ spanning $z=[2,4]$), considering the sparse distribution of quasars at $z>4$, 30 $k$-bins ($\Delta k=0.01$ Mpc$^{-1}$) from 0.06 to 0.35 Mpc$^{-1}$, and 4 $\mu$-bins ($\Delta \mu=0.25$) covering $0<\mu<1$. 

For 21~cm IM, we consider the low-frequency array of the Square Kilometre Array in Phase 1 (SKA1-Low), and Stage II IM surveys, e.g., Packed Ultra-wideband Mapping Array (PUMA, \citealt{2018arXiv181009572C}). The total observed 21\,cm power spectrum includes thermal noise from radio interferometer,
\begin{eqnarray}
    \label{eq:P_21_tot}
    P_\mathrm{21, Tot}^\mathrm{fid}(\boldsymbol{k},z) = P_{21}^\mathrm{fid}(\boldsymbol{k},z) + P_\mathrm{N} (\boldsymbol{k},z) \, .
\end{eqnarray}
$P_\mathrm{N}$ is given by \citep{2015ApJ...803...21B,2018JCAP...05..004O}
\begin{eqnarray}
    P_\mathrm{N} (\boldsymbol{k},z) & = & T^2_\mathrm{sys}(z) \chi^2(z) \lambda (z) \frac{1 + z}{H(z)} \left(\frac{\lambda^2 (z)}{A_e}\right)^2 \left(\frac{S_\mathrm{area}}{\mathrm{FOV}(z)}\right) \nonumber \\
    \label{eq:p_thermal}
    & & \times \left(\frac{1}{N_\mathrm{pol} t_\mathrm{int} n_\mathrm{b}(u = k_{\perp} \chi (z)/ 2\pi)}\right) \, ,
\end{eqnarray}
where $T_\mathrm{sys}$ is the system temperature of the instrument \citep{2015JCAP...03..034V, 2018arXiv181009572C}:
\begin{equation}
T_{\rm sys}^{\rm SKA1\mbox{-}Low}(\nu)
=60\left(\frac{300\,{\rm MHz}}{\nu}\right)^{2.55}
\times 1.1 + 40 ,
\end{equation}
\begin{equation}
\begin{aligned}
T_{\rm sys}^{\rm Stage\,II}(\nu)
=&\,
25\left(\frac{400\,{\rm MHz}}{\nu}\right)^{2.75}
+ 2.7
\\
&+
\frac{300}{9}
+
\frac{50}{0.81} .
\end{aligned}
\end{equation}
$\chi(z)$ is the comoving distance; $\lambda(z)=\lambda_{21}(1+z)$ is the observed 21\,cm wavelength. $A_e$ denotes the effective collecting area. For Stage II IM, $A_e$ scales with dish area with 70\% aperture efficiency and $D_\mathrm{phys}=6\,$m, thus $A_e=0.7\pi (D_\mathrm{phys}/2)^2$. For SKA1-Low, stations include 256 antennas, with collecting area decreasing as $\nu^{-2}$ above a critical frequency and constant below \citep{2015aska.confE..19S,2020PASA...37....7S}, i.e.,
\begin{align}
\begin{split}
    A_e(\nu)=\left\{
    \begin{array}{lc}
    3.2\times256, & \nu\leq\nu_\mathrm{crit}  \\
    3.2\times256\times(\nu_\mathrm{crit}/\nu)^2, & \nu>\nu_\mathrm{crit}
    \end{array}
    \right .
\end{split}
\end{align}
where $\nu_\mathrm{crit}=110\,\mathrm{MHz}$.

Moreover, $S_\mathrm{area}$ is the survey area, with field of view $\mathrm{FOV}(z)=(\lambda(z)/D_\mathrm{eff})^2$, where $D_\mathrm{eff}=\sqrt{0.7}D_\mathrm{phys}$. For SKA1-Low, $D_\mathrm{phys} = 40\,\mathrm{m}$ and we assume a deep-narrow survey, i.e. $S_\mathrm{area}=\mathrm{FOV}(z)$ to reduce thermal noise. Stage II IM will observe half the sky. $N_\mathrm{pol}=2$ is the number of polarizations, $t_\mathrm{int}$ is the total integration time, and $n_\mathrm{b}$ denotes the baseline density in the $uv$-plane, averaged over 24h for sky rotation. For SKA1-Low, we assume $t_\mathrm{int}=5000\,$h and $n_\mathrm{b}$ from Figure~6 of  \citet{2015JCAP...03..034V} normalized to match baseline counts
\begin{equation}
    \int 2\pi un_\mathrm{b}(u)du=\frac{N_\mathrm{b}(N_\mathrm{b}-1)}{2},
\end{equation}
where $N_\mathrm{b}=224$ since we only consider stations in the dense core of the array. For Stage II IM, we assume $t_\mathrm{int}=1000\,$h. $n_\mathrm{b}$ is calculated via hexagonal close-packing  in a compact circle \citep{2018arXiv181009572C}, normalized with $N_\mathrm{b}=32000$. 

The surveys use different binning:\begin{itemize}
    \item SKA1-Low: 7 redshift bins ($3.5<z<5.5$, $\Delta z=0.3$), 5 $\mu$-bins ($0<\mu<1$, $\Delta \mu=0.2$). We optimistically assume the wedge can be cleaned to reduce the covariance.
    \item Stage II IM: 10 redshift bins ($3.5<z<5.5$, $\Delta z=0.2$). We conservatively retain the wedge and use 5 linearly-spaced $\mu$-bins ($\mu_\mathrm{min}<\mu<1$), with  \citep{2016MNRAS.456.3142S}
\end{itemize}
\begin{equation}
    \mu_\mathrm{min}=\frac{\chi(z)H(z)/[c(1+z)]}{\sqrt{1+\{\chi(z)H(z)/[c(1+z)]\}^2}}\, .
\end{equation}

We follow previous work \citep{2023MNRAS.525.6036L} for the choice of the $k$-bins: 30 logarithmically spaced $k$-bins with $k_\mathrm{max}=0.4 \ \mathrm{Mpc^{-1}}$ and $k_\mathrm{min}=\sqrt{k_\mathrm{\parallel,min}^2+k_\mathrm{\perp,min}^2}$, where $k_\mathrm{\parallel,min}$ is set by the depth of the given redshift bin 
\begin{equation}
    \label{eq:km_par}
    k_\mathrm{\parallel,min}=\frac{2\pi}{\chi(z_\mathrm{max})-\chi(z_\mathrm{min})}\, ,
\end{equation}
and $k_\mathrm{\perp,min}$ is limited by the shortest baseline 
\begin{equation}
    \label{eq:km_per}
    k_\mathrm{\perp,min}=\frac{2\pi D_\mathrm{phys}}{\lambda(z)\chi(z)}\, .
\end{equation}
The bins simultaneously satisfy $k\times\mu\geq k_\mathrm{\parallel,min}$ and $k\times \sqrt{1-\mu^2} \geq k_\mathrm{\perp,min}$.

\subsection{Results}
\label{sec:mcmc_results}
\begin{figure}[htb!]
\centering
\includegraphics[width=\columnwidth]{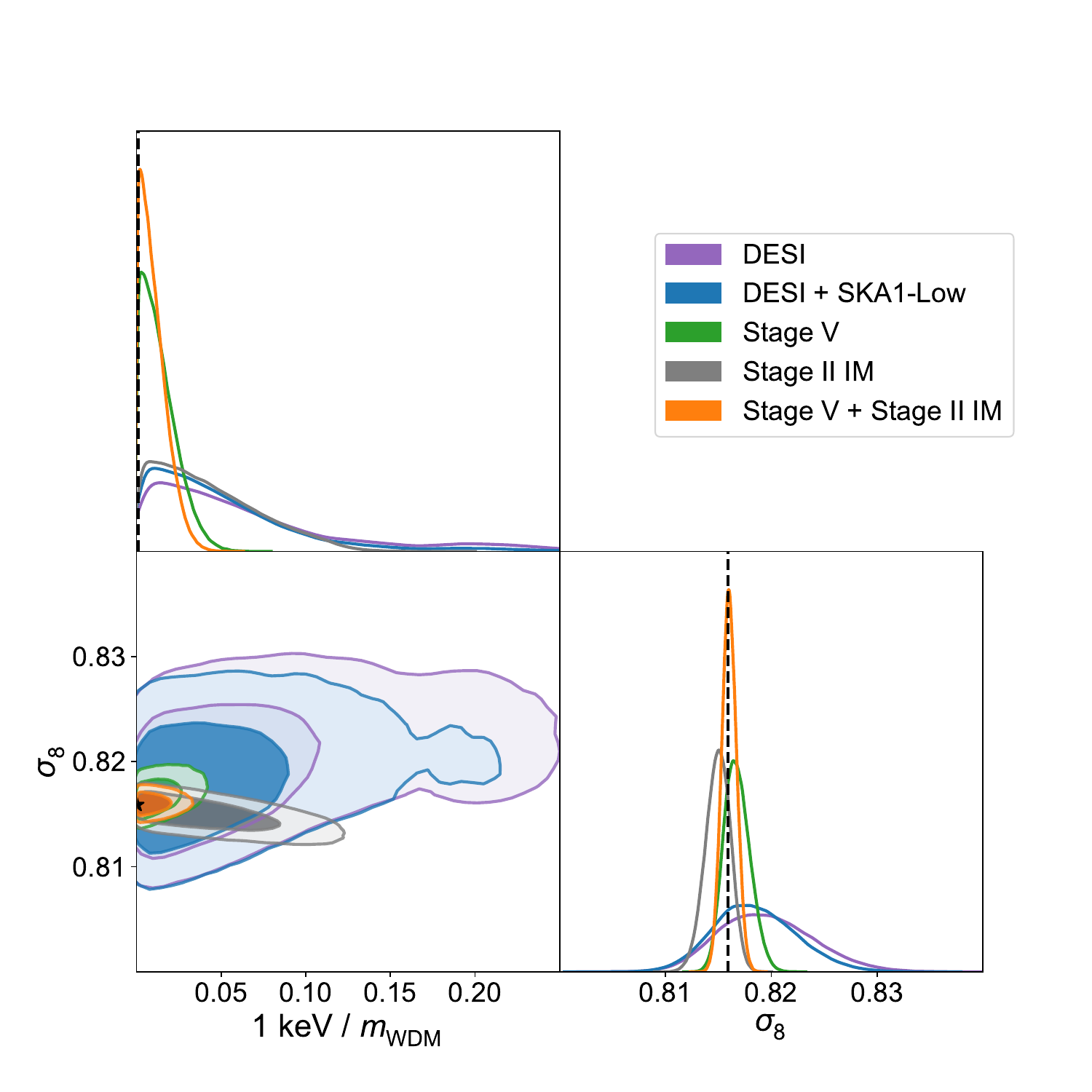}
\caption{Forecast of the constraints on $m_\mathrm{WDM}$ and $\sigma_8$.
We show the constraints by DESI-like \lya survey (purple), the combination of DESI and SKA1-Low 5,000-h observation of 21 cm IM (blue), Stage V \lya surveys (green), Stage II IM 1,000-h observation (grey), and the combination of Stage V \lya surveys and Stage II IM (orange), respectively, with the dark (light) contours representing the 68\% (95\%) credible regions. Our fiducial model is the CDM, $1\ \mathrm{keV}/m_\mathrm{WDM}=0$ and $\sigma_8=0.8159$ (black star).}
\label{fig:mcmc}
\end{figure}

The MCMC forecast results are shown in Figure~\ref{fig:mcmc}, and all constraints quoted below correspond to 95\% credible intervals. Using the large-scale \lya forest power spectrum including reionization relics from a DESI-like survey, we obtain a constraint of $m_\mathrm{WDM}>5.0$ keV. This is comparable to current constraints derived from small-scale \lya forest observations \citep{2017PhRvD..96b3522I,2020JCAP...04..038P,2024PhRvD.109d3511I} (the strongest is $m_\mathrm{WDM}>5.7\,\mathrm{keV}$ given by \citealt {2024PhRvD.109d3511I}; see Figure~\ref{fig:mass_wdm_constraints}).

We find that the constraining power of the 21~cm IM power spectrum from SKA1-Low alone is weak because of the sparse baseline density and the resulting large instrumental noise. Nevertheless, combining the \lya forest with 21~cm IM improves the constraint by breaking parameter degeneracies through their distinct redshift-dependent responses to $m_\mathrm{WDM}$. In particular, the joint analysis strengthens the constraint to $m_\mathrm{WDM}>7.1\,\mathrm{keV}$, making it among the most competitive constraints currently achievable.

For next-generation surveys, large-scale \lya forest power spectrum measurements from Stage-V spectroscopic surveys yield an exceptionally strong constraint of $m_\mathrm{WDM}>30\,\mathrm{keV}$. Meanwhile, 21~cm IM observations with Stage II experiment independently achieve $m_\mathrm{WDM}>10\,\mathrm{keV}$, representing a substantial improvement over SKA1-Low. This improvement is primarily driven by the significantly increased baseline density, with the Stage II array containing roughly four orders of magnitude more baselines than the 224 core stations of SKA1-Low. More importantly, the combination of Stage-V \lya forest and Stage II 21~cm IM further breaks the degeneracy between $m_\mathrm{WDM}$ and $\sigma_8$, leading to a constraint of $m_\mathrm{WDM}>39\,\mathrm{keV}$. We emphasize that these forecasts do not include the \lya forest--21~cm cross-correlation signal \citep{2025MNRAS.536.1645M}, which could potentially further improve the constraints.

\subsection{Impact of reionization astrophysics}
Our results are robust even if we include the astrophysical parameters ionization efficiency $\zeta$ and turnover mass of halos unable to host star-forming galaxies $M_\mathrm{turn}$, which introduce astrophysics uncertainties (see Appendix~\ref{app:4d-mcmc} for details of the method). $\zeta$ governs ionizing photon escape and reionization timing, while $M_\mathrm{turn}$ primarily sets the halo population contributing photons, shaping reionization onset, pace, and morphology. As WDM suppresses small-scale structure, it changes the timeline of reionization; however, this change can be mimicked by reionization astrophysics, as shown in Figure~\ref{fig:global_history}. Besides, WDM can also accelerate the late stages of reionization \citep{2014MNRAS.438.2664S}, therefore we require at least both $\zeta$ and $M_\mathrm{turn}$ to mimic these trends. Modest variations in $\zeta$ and $M_\mathrm{turn}$ relax the $m_\mathrm{WDM}$ constraints from DESI, DESI + SKA1-Low, Stage V \lya surveys, and Stage V + Stage II IM by 12\%, 12\%, 40\% and 36\%, respectively (see  Table~\ref{tab:mcmc_comparison} for details). Even with these astrophysics uncertainties, the constraints from next-generation surveys remain capable of pushing well beyond current bounds.

\begin{figure*}[htb!]%
\centering
\includegraphics[width=1.0\textwidth]{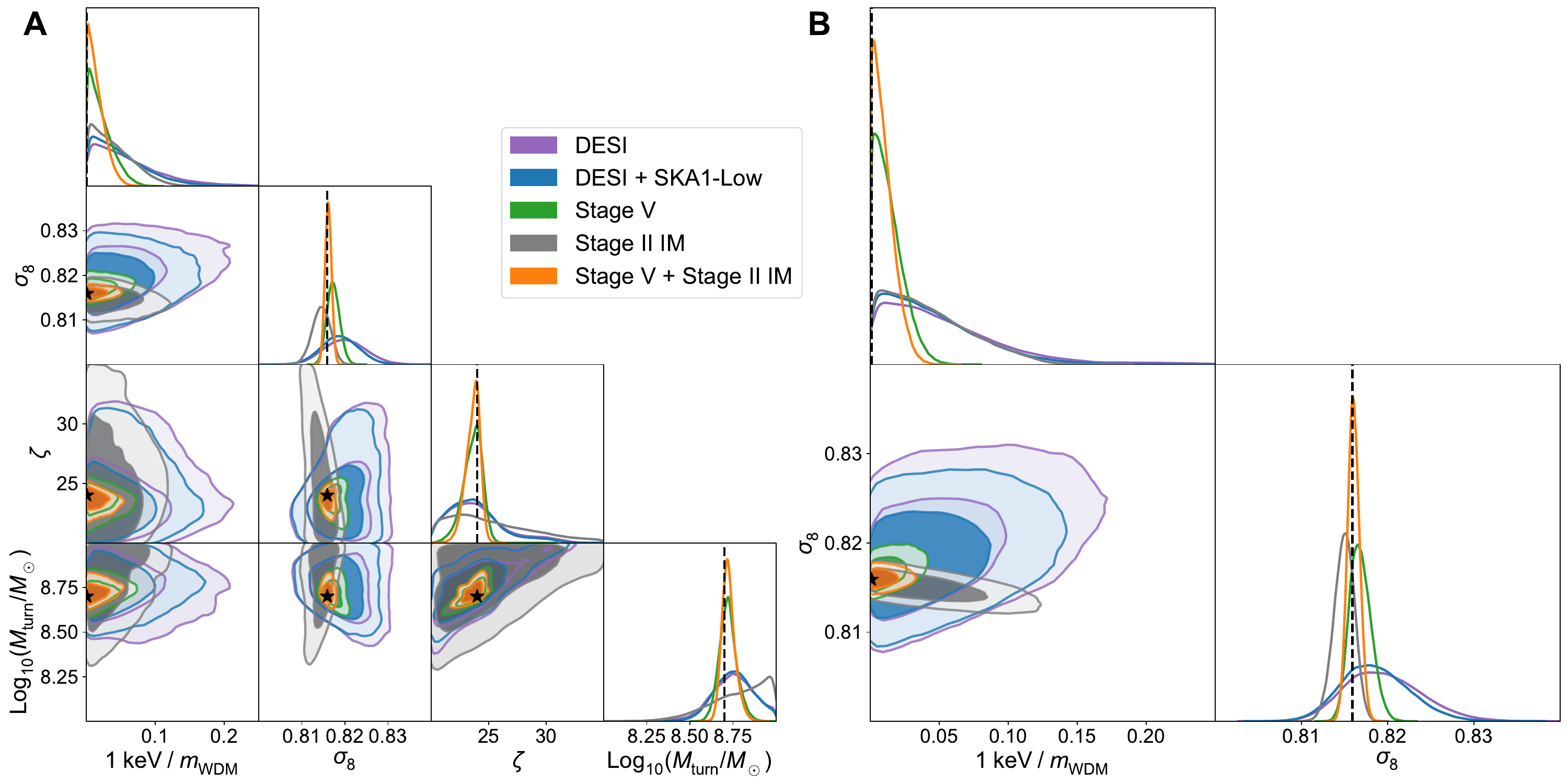}
\caption{Comparison of 4D and 2D MCMC forecasts using a single realization. (A) The 4D MCMC introduces two additional free parameters, ionizing efficiency $\zeta$ and turnover mass of halos unable to host star-forming galaxies $M_\mathrm{turn}$, which represent the uncertainties of reionization astrophysics. The fiducial model is CDM ($1\ \mathrm{keV}/m_\mathrm{WDM}=0$), $\sigma_8=0.8159$, $\zeta=24$ and $\mathrm{Log_{10}}(M_\mathrm{turn}/M_\odot)=8.7$. (B) The 2D MCMC fixes $\zeta$ and $M_\mathrm{turn}$ at the fiducial values. The conventions are the same as Figure \ref{fig:mcmc}.}
\label{fig:2D_4D}
\end{figure*}

More interestingly, Figure~\ref{fig:2D_4D} shows that the degeneracy between $\zeta$ and $M_\mathrm{turn}$ with $m_\mathrm{WDM}$, which is present in $x_\mathrm{HI}(z)$, is gone in our constraints by post-reionization observables. Figure~7 and 8 of \citet{2021MNRAS.508.1262M}, demonstrate that an increase in $\zeta$ or a reduction in $M_\mathrm{turn}$ both yield a decrease in the resulting 3D flux power spectrum, since both changes make reionization occur earlier, which diminishes the large scale enhancement cause by reionization relics. As shown in Figure~\ref{fig:power_spectrum}, increasing $m_\mathrm{WDM}$ also reduces the 3D flux power spectrum at high redshifts. However, at low redshifts, increasing $m_\mathrm{WDM}$ leads to an \emph{increase} in the flux power spectrum. This redshift-denpendent effect of WDM breaks the degeneracy between reionization astrophysics and $m_\mathrm{WDM}$ in our signal. This interesting feature means that -- ignoring the instrument systematics and signal-to-noise -- even if higher redshifts translate to better sensitivity to the impact of warm dark matter in the \lya forest, incorporating lower redshift observations would be crucial to disentangle its effects from those of reionization astrophysics. 

\begin{table*}[htb!] 
\centering
\caption{Comparison between 4D (cosmology + reionization astrophysics) and 2D (cosmology only)
MCMC forecasts using the same single realization.
The table reports 95\% credible intervals for $m_\mathrm{DM}$, $\zeta$, and
$\log_{10}(M_\mathrm{turn}/M_\odot)$, and 68\% credible intervals for $\sigma_8$.
The fiducial model assumes CDM ($m_\mathrm{DM}=\infty$), $\sigma_8=0.8159$,
$\zeta=24$, and $\log_{10}(M_\mathrm{turn}/M_\odot)=8.7$.
Note that the 2D MCMC constraints here differ from the results in Section~\ref{sec:mcmc_results} and Figure \ref{fig:mcmc}, as the latter are based on the average of four realizations, whereas the former uses only one.}
\label{tab:mcmc_comparison}

\begin{tabular}{lccccc}
\\
\hline
& DESI & DESI+SKA1-Low & Stage V & Stage II IM & Stage II IM+Stage V \\
\hline
\multicolumn{6}{c}{$m_\mathrm{DM}$ (keV, 95\% credible interval)} \\
\hline
2D & $>7.3$ & $>8.9$ & $>30$ & $>10.0$ & $>39$ \\
4D & $>6.4$ & $>7.8$ & $>18$ & $>10.5$ & $>25$ \\
\hline
\multicolumn{6}{c}{$\sigma_8$ (68\% credible interval)} \\
\hline
2D & $0.8192^{+0.0047}_{-0.0041}$ & $0.8183^{+0.0042}_{-0.0036}$ &
     $0.8168^{+0.0013}_{-0.0011}$ & $0.8150^{+0.0011}_{-0.0012}$ & $0.8160\pm0.0007$ \\
4D & $0.8199^{+0.0050}_{-0.0048}$ & $0.8188^{+0.0043}_{-0.0042}$ &
     $0.8172^{+0.0015}_{-0.0014}$ & $0.8145^{+0.0020}_{-0.0021}$ &
     $0.8161^{+0.0008}_{-0.0007}$ \\
\hline
\multicolumn{6}{c}{$\zeta$ (95\% credible interval)} \\
\hline
4D & $23.6^{+6.8}_{-3.3}$ & $23.5^{+6.3}_{-3.2}$ &
     $23.7^{+1.3}_{-2.1}$ & $24.5^{+9.0}_{-4.2}$ & $23.7^{+1.0}_{-1.5}$ \\
\hline
\multicolumn{6}{c}{$\log_{10}(M_\mathrm{turn}/M_\odot)$ (95\% credible interval)} \\
\hline
4D & $8.75^{+0.22}_{-0.26}$ & $8.76^{+0.21}_{-0.22}$ &
     $8.72^{+0.11}_{-0.10}$ & $8.81^{+0.18}_{-0.40}$ & $8.72^{+0.09}_{-0.06}$ \\
\hline
\end{tabular}
\end{table*}

\section{Discussion}
\label{sec:disscusion}
It is well established that the IGM's large-scale fluctuations of temperature and ionization following patchy reionization boost the large-scale \lya forest power spectrum. Our methodology gives 10-20\% 1D power boost at $k=0.14\; \mathrm{Mpc^{-1}}$ and $z=4$ in CDM \citep{2020MNRAS.499.1640M}, comparable with 5-20\% at $z=4.2$ in hybrid hydrodynamical and radiation transfer simulations \citep{2022MNRAS.509.6119M} and 10-30\% given by hybrid semi-numerical reionization model and hydrodynamical simulations \citep{2019MNRAS.486.4075O}, while  \citep{2019MNRAS.490.3177W} reports 20-40\% enhancement at $z=5$. While some self-consistent simulations couple patchy reionization and inhomogeneous heating \citep{2019MNRAS.486.4075O,2019MNRAS.490.3177W,2022MNRAS.509.6119M,2023MNRAS.519.6162P}, their gas resolution ($1\times10^5 - 2\times10^6 M_\odot$) cannot resolve gas below the pre-reionization Jeans mass, therefore is insufficient to capture the long-term survival of reionization relics \citep{2018MNRAS.474.2173H,2024MNRAS.533L.100C}\footnote{See \citealt{2026arXiv260325788C} for recent simulation efforts along this direction.}. We identify WDM-dependent, long-lasting reionization relics at $z<4$, enabled by the high mass resolution that resolves the hydrodynamical responses of mini-halos, filaments, and voids after reionization, and the evolution of HEMD gas. In our hybrid method the small-box hydrodynamical simulations are coupled to the \textsc{21cmFAST} simulations via reionization redshifts (Equation \ref{eq:cross_power}), implicitly coupling them to different density environments, as overdense regions tend to reionize earlier, although the small-box total density is fixed at the cosmic mean. This hybrid methodology inevitably leads to an underestimation of mode coupling between large and small-scale modes \citep{2019MNRAS.487.1047M}. 

Current measurement's uncertainties of \lya forest monopole and quadrupole at $z=2.3$ are 3-7\% \citep{2024MNRAS.533.3756D}, exceeding the WDM-CDM differences predicted by our effect at $z=2.3$ ($\lesssim1\%$). Thus, current multipole data is not suitable for this program. However, a similar-precision future measurement at $z=3$ would be viable, as the predicted differences reach comparable levels with the measurement's uncertainty. Furthermore, the detection of 21 cm emission at $z=2.3$ in cross-correlation with \lya forest marks the beginning of 21 cm IM analyses at $z>2$ \citep{2024ApJ...963...23A}. Following  \cite{2025MNRAS.536.1645M}, we calculate the cross-power spectrum of \lya forest and 21 cm IM including reionization relics. The inclusion increases the signals and the lighter WDM has stronger effect. At $k=0.05\;\mathrm{Mpc^{-1}}$, the differences of the LoS power between 3 keV WDM and CDM are 3.6\% and 9.3\% at $z=3.5$ and 4, respectively.

\section{Conclusion}
\label{sec:conclusion}

We introduce a new large-scale probe of WDM based on the long-lived ``memory of reionization'' in post-reionization observables. Although WDM primarily suppresses structure formation below the free-streaming scale, reionization couples this small-scale physics to large scales: the \emph{local} reionization redshift, $z_\mathrm{re}$, is spatially inhomogeneous on ionized-bubble scales (several to tens of comoving Mpc), and post-reionization gas properties retain a measurable and WDM-dependent memory of $z_\mathrm{re}$ down to $z\sim2$--$5$. Consequently, the spatial variations in $z_{\rm re}$ modulate both the Ly$\alpha$ forest transparency and the halo \HI\ content traced by 21\,cm IM, introducing large-scale sensitivity to WDM physics.

For the Ly$\alpha$ forest, patchy reionization enhances large-scale power because overdense regions typically reionize earlier and have more time to cool, increasing their recombination rate and opacity. This amplifies large-scale transparency contrast on ionized-bubble scales. The level of this enhancement is WDM-dependent for two reasons: (1) WDM alters the post-reionization thermal and dynamical response of small-scale structures (including the production and evolution of HEMD gas), which changes how strongly IGM transparency depends on $z_\mathrm{re}$; and (2) WDM tends to delay reionization by suppressing low-mass sources, increasing the fraction of regions that reionize late and therefore retain stronger reionization memory at a fixed $z_\mathrm{obs}$. Quantitatively, including reionization relics, the fractional difference between WDM and CDM Ly$\alpha$ power at $z=4$ reaches $\sim$19\% for $m_\mathrm{WDM}=3$\,keV and $\sim$2\% for $m_\mathrm{WDM}=9$\,keV at $k=0.05\,{\rm Mpc}^{-1}$. Without reionization relics, these large-scale differences largely vanish.

For 21\,cm IM, reionization relics suppress large-scale power because photoheating reduces the baryonic content (and hence the \HI\ mass) of low-mass halos more strongly in regions that reionize earlier. When smoothed on ionized-bubble scales, this stronger reduction in overdense regions decreases \HI\ fluctuations and therefore suppresses 21\,cm brightness-temperature fluctuations. The suppression is weaker for lighter WDM because fewer low-mass halos exist to be affected and the sensitivity of \HI\ density to $z_\mathrm{re}$ is reduced. The WDM-CDM fractional differences at $z=5.5$ are $\sim$19\% for $m_\mathrm{WDM}=3$\,keV and and $\sim$5\% for $m_\mathrm{WDM}=9$\,keV at $k=0.05\,{\rm Mpc}^{-1}$.

Using these WDM-dependent large-scale signatures, we performe MCMC forecasts for constraints on $m_\mathrm{WDM}$ jointly with $\sigma_8$. For a Ly$\alpha$ survey with a covariance matrix designed to mimic DESI-like performance, we forecast a lower bound of
$m_\mathrm{WDM} > 5.0~\mathrm{keV}\  (95\%)$. Adding 21\,cm IM from a 5,000\,h SKA1-Low-like survey breaks degeneracies and strengthens the constraint to $m_\mathrm{WDM} > 7.1~\mathrm{keV}$. Looking further ahead, we find that next-generation experiments can be substantially more powerful: a Stage-V Ly$\alpha$ survey yields $m_\mathrm{WDM}>30$\,keV, a Stage-II IM survey yields $m_\mathrm{WDM}>10$\,keV, and their combination yields
$m_\mathrm{WDM} > 39~\mathrm{keV}$. Our forecasts demonstrate that large-scale measurements can give competitive (and potentially leading) constraints on WDM when considering reionization relics.

We also assess robustness to uncertainties in reionization astrophysics by allowing the ionizing efficiency $\zeta$ and the turnover mass $M_\mathrm{turn}$ to vary. These parameters can partially mimic WDM in the global reionization history; however, the power spectra retain additional discriminatory power because the WDM imprint is mediated by small-scale gas dynamics and exhibits a characteristic redshift dependence that differs from simple shifts in reionization timing.

More broadly, our results highlight reionization relics as a generic pathway to connect small-scale dark-matter physics to large-scale post-reionization observables. Our method can be applied to other dark-matter candidates, likely with similarly tight constraints, such as fuzzy dark matter \citep[e.g.,][]{2000PhRvL..85.1158H}, primordial black holes \citep[e.g.,][]{2019JCAP...08..031M}, and self-interacting dark matter \citep[e.g.,][]{2000PhRvL..84.3760S}, provided that inhomogeneous reionization and the post-reionization evolution of gas on small scales can be modeled accurately. Reionization relics therefore offer a complementary and potentially powerful probe of the nature of dark matter.

\begin{acknowledgments}
This work is supported by the National SKA Program of China (grant No. 2020SKA0110401), National Natural Science Foundation of China (grant No. 11821303), the Basic and Frontier Research Project of PCL (grant No. 2025QYB012), and the Major Key Project of PCL. The authors acknowledge the Tsinghua Astrophysics High-Performance Computing platform at Tsinghua University and PCL's Cloud Brain for providing computational and data storage resources
that have contributed to the research results reported within this paper.
\end{acknowledgments}

\begin{contribution}

P.M.C.\ and Y.M.\ initiated and oversaw the project. P.M.C. contributed to the forecast, analysis, and setup of both small-box and large-box simulations. Y.Z.\ led the analysis of the small-box simulations, ran the small-box and large-box simulations, performed the forecast, and led the drafting of the paper. C.M.G.\ led the analysis of the large-box simulations, ran the large-box simulations, and contributed extensively to figures. H.L. contributed to the analysis of the small-box snapshots to compute the \HI\ IM signal. C.M.H.\ and Y.M.\ contributed extensively to the analysis, discussion, and interpretation of the results. All authors contributed to the text, figures, editing, and formatting.

\end{contribution}

\software{classy \citep{2011JCAP...07..034B}, 
          emcee \citep{2013PASP..125..306F}, 
          21cmFAST \citep{2011MNRAS.411..955M,2020JOSS....5.2582M,2019MNRAS.484..933P}, 
          Gadget-2 \citep{2001NewA....6...79S, 2005MNRAS.364.1105S}, 
          ncdm-mass-functions \citep{2022MNRAS.509.1703S}
          }

\section*{Data Availability}
The data generated in this work, including IGM transparency, \HI\ density, the cross-power spectrum of matter density and neutral hydrogen fraction, the evolution of global neutral hydrogen fraction, the \lya forest and 21 cm IM power spectra, and the custom code for data analysis and forecast used in this work are available at  \url{https://github.com/yaozhang1206/WDM_Lya_21cm}. There are no restrictions on data availability.

\appendix

\section{WDM constraints}
\label{App:WDM}

Figure~\ref{fig:mass_wdm_constraints} summarizes current $m_\mathrm{WDM}$ constraints from \lya forest measurements and 21 cm IM projections. Several \lya constraints use Sloan Digital Sky Survey (SDSS) \citep{2000AJ....120.1579Y} observations, including \citet{2006PhRvL..97g1301V,2008PhRvL.100d1304V,2009JCAP...05..012B}, as well as analysis incorporating High-Resolution Echelle Spectrometer (HIRES) \citep{1994SPIE.2198..362V} data, e.g., \cite{2008PhRvL.100d1304V,2013PhRvD..88d3502V,2017PhRvD..96b3522I,2017JCAP...06..047Y,2017PhLB..773..258G,2021MNRAS.502.2356G,2023PhRvD.108b3502V,2024PhRvD.109d3511I}, and constraints using Wilkinson Microwave Anisotropy Probe (WMAP5) \citep{2009ApJS..180..306D} five-year data \citep{2009JCAP...05..012B}. Additional constraints use Magellan Inamori Kyocera Echelle spectrograph (MIKE) \citep{2002SPIE.4485..453B} observations, e.g., \cite{2013PhRvD..88d3502V,2017PhRvD..96b3522I,2017JCAP...06..047Y,2017PhLB..773..258G}, Baryon Oscillation Spectroscopic Survey (BOSS/SDSS-III) data \citep{2016JCAP...08..012B,2017JCAP...06..047Y}, XQ-100 \citep{2017PhRvD..96b3522I,2017JCAP...06..047Y,2020JCAP...04..038P}, extended Baryon Oscillation Spectroscopic Survey (eBOSS) data \citep{2020JCAP...04..038P}, and Ultraviolet and Visual Echelle Spectrograph (UVES) data \citep{2000SPIE.4008..534D}, e.g., \cite{2024PhRvD.109d3511I,2023PhRvD.108b3502V}. \cite{2024PhRvD.109d3511I} examined patchy reionization's impact on WDM constraints from small-scale \lya forest, finding a 10\% weakening of the constraint. Their simulations lack the mass resolution to resolve the HEMD gas studied here, and their patchy reionization correction was derived solely using CDM, neglecting the WDM influence on such corrections. 

For 21 cm IM, constraints are based on data from the Experiment to Detect the Global EoR Signature (EDGES) \citep{2018Natur.555...67B}, e.g., \citet{2018ApJ...859L..18S,2019MNRAS.487.3560C,2019PhRvD.100l3005B} and forecasts of EDGES-like instruments, e.g., \cite{2020PhRvD.101f3526M}, or forecasts for the low-frequency array of the Square Kilometre Array in Phase 1 (SKA1-Low) (SKA1-Low) \citep{2019arXiv191212699B}, e.g., \cite{2015JCAP...07..047C,2020PASA...37....7S,2023JCAP...03..047M}, and for the Hydrogen Epoch of Reionization Array (HERA) \citep{2017PASP..129d5001D}, e.g., \cite{2020PhRvD.101f3526M}.

\section{Impact of reionization astrophysics (4-parameter MCMC)}
\label{app:4d-mcmc}

To explore the impact of reionization astrophysics on $m_\mathrm{WDM}$ constraints, we vary the ionizing efficiency $\zeta$ and turnover mass $M_\mathrm{turn}$ in \textsc{21cmFAST}. Our 2-parameter MCMC has $\zeta=24$ and $M_\mathrm{turn}=5\times10^8\; M_\odot$. Here we extend the simulation grid with 24 single-realization runs for $m_\mathrm{DM}=\{3\,\mathrm{keV},\ \mathrm{\infty(CDM)}\}$, $\sigma_8=\{0.7659,\ 0.8659\}$, $\zeta=\{20, 35\}$, and $M_\mathrm{turn}=\{10^8,\ 5\times10^8,\ 10^9\}M_\odot$, plus 15 simulations from the 2-parameter MCMC with the same seed. 

The 2-parameter MCMC forecast in Figure~\ref{fig:mcmc} averages four realizations of \lya forest and 21 cm IM power spectra. In contrast, the 4-parameter MCMC uses a single realization, making direct comparison inappropriate. For a valid comparison, we also conduct a 2-parameter MCMC on the same realization (see Figure~\ref{fig:2D_4D} and Table~\ref{tab:mcmc_comparison}).

Figure~\ref{fig:2D_4D} shows Stage II IM constraints (grey) are bounded by $\zeta=20$ and $\log_{10}(M_\mathrm{turn}/M_\odot)=9$, suggesting a 5\% tighter $m_\mathrm{WDM}$ limit than the 2-parameter MCMC. Nevertheless, Figure~\ref{fig:global_history} reveals some models with $\zeta<20$ and $\log_{10}(M_\mathrm{turn}/M_\odot)>9$ (e.g $m_\mathrm{WDM}=3$ keV, $\sigma_8=0.7659$, the leftmost dashed line) reionize too late to match $x_\mathrm{HI}$ observations, while others (e.g., CDM with $\sigma_8=0.8659$) allow for more delayed reionization. To avoid \emph{extremely} late reionization, $\zeta$ lower and $M_\mathrm{turn}$ upper limits should depend on $m_\mathrm{WDM}$ and $\sigma_8$, creating a ``wedge-like'' prior. Extended parameter space and prior impact exploration are deferred to future work.

The small-box simulations use a fixed photoheating rate after reionization. We test variations using half/double photoheating rates. Higher heating reduces \lya forest reionization relics (weaker IGM cooling and weakened dependence of $\psi$ on $z_\mathrm{re}$) but amplifies 21 cm IM relics (stronger baryon suppression in low-mass halos). We do not consider \HeII\ reionization ($z_\mathrm{re}^\mathrm{He_{II}}\approx3$) \citep{2021MNRAS.506.4389G}. Its heat injection likely reduces sensitivity of IGM transparency to $z_\mathrm{re}$ \citep{2018MNRAS.474.2173H} while corresponding temperature fluctuations can enhance the \lya power spectrum \citep{2024MNRAS.535.1035M}. We leave these effects to future work.

\bibliography{ref}{}
\bibliographystyle{aasjournalv7}

\end{document}